\documentclass[twocolumn]{aastex631}

\graphicspath{{./}{figures/}}
\usepackage{mwe}
\usepackage[position = top]{subfig}
\usepackage{amsmath}
\DeclareMathOperator{\sech}{sech} 
\usepackage{appendix}
\usepackage{graphicx}
\usepackage{mathtools}
\usepackage{diagbox}
\usepackage{footmisc}
\usepackage{multirow}
\def\code#1{\texttt{#1}}

\begin{document}

\title{Particle Injection and Nonthermal Particle Acceleration in Relativistic Magnetic Reconnection\footnote{Released \today}}

\author[0000-0001-6155-2827]{Omar French}\footnote{omar.french@colorado.edu}
\affiliation{Center for Integrated Plasma Studies, Department of Physics, 390 UCB, University of Colorado, Boulder, CO 80309, USA}
\affiliation{Los Alamos National Laboratory, Los Alamos, NM 87545, USA}

\author[0000-0003-4315-3755]{Fan Guo}
\affiliation{Los Alamos National Laboratory, Los Alamos, NM 87545, USA}

\author[0000-0002-9504-641X]{Qile Zhang}
\affiliation{Los Alamos National Laboratory, Los Alamos, NM 87545, USA}

\author[0000-0001-8792-6698]{Dmitri A. Uzdensky}
\affiliation{Center for Integrated Plasma Studies, Department of Physics, 390 UCB, University of Colorado, Boulder, CO 80309, USA}

\begin{abstract}

Magnetic reconnection in the relativistic regime has been proposed as an important process for the efficient production of nonthermal particles and high-energy emissions. Using fully kinetic particle-in-cell simulations, we investigate how guide-field strength and domain size affect characteristic spectral features and acceleration processes. We study two stages of acceleration: energization up until the injection energy~$\gamma_{\rm inj}$ and further acceleration that generates a power-law spectrum. Stronger guide fields increase the power-law index and~$\gamma_{\rm inj}$, which suppresses acceleration efficiency. These quantities seemingly converge with increasing domain size, suggesting that our findings can be extended to large-scale systems. We find that three distinct mechanisms contribute to acceleration during injection: particle streaming along the parallel electric field, Fermi reflection, and the pickup process. Fermi and pickup processes, related to the electric field perpendicular to the magnetic field, govern the injection for weak guide fields and larger domains. Meanwhile, parallel electric fields are important for injection in the strong guide field regime. In the post-injection stage, we find that perpendicular electric fields dominate particle acceleration in the weak guide field regime, whereas parallel electric fields control acceleration for strong guide fields. These findings will help explain the nonthermal acceleration and emissions in high-energy astrophysics, including black hole jets and pulsar wind nebulae.

\end{abstract}

\section{Introduction} \label{sec:1_introduction}

A fundamental task in high-energy astrophysics is understanding how some particles obtain a large amount of energy and radiate it away. Energetic particles that gain energy far beyond the average particle energy can generate high-energy emissions via different radiation processes. In the study of astrophysical jets from active galactic nuclei (AGN), for example, the origin of energetic particles and the source of high-energy emission is a topic of intensive debate \citep{Romanova1992, Giannios2009,Giannios2010, Sironi2015,Blandford2019,Zhang2020}. In pulsar wind nebulae (PWNe), several distinct regions can contribute to the overall particle acceleration \citep{Rees1974,Uzdensky2011b,Sironi2011,Komissarov2011,Cerutti2013,Sironi2017,Lyutikov2018,Cerutti2020,Lu2021}. The origin of high-energy emission often involves particles that are distributed in the form of nonthermal power-law energy spectra. Consequently, a detailed description of nonthermal particle acceleration is essential for understanding high-energy emissions.

One of the main proposed mechanisms for nonthermal particle acceleration in space and astrophysical plasmas is magnetic reconnection, a process that rearranges magnetic topology (see Figure~\ref{fig:accel_cartoons}a for an illustration) and rapidly liberates magnetic energy into heat, bulk flows, and the acceleration of nonthermal particles \citep{Biskamp2000,Zweibel2009,Yamada2010,Ji2022,Yamada2022}.

Fully kinetic particle-in-cell (PIC) simulations of collisionless magnetic reconnection enable studies of nonthermal particle acceleration directly from first principles. In the relativistic regime relevant to AGN jets and PWNe, the upstream ambient ``hot" magnetization parameter~$\sigma_h \equiv B_0^2/4\pi h$ (i.e., the enthalpy density of the reconnecting magnetic field divided by the relativistic enthalpy density~$h$ of the upstream plasma) is often very large ($\sigma_h \gg 1$), leading to strong nonthermal particle acceleration \citep[see][for focused reviews]{Hoshino2012,Kagan2015,Guo2020}. Numerous PIC simulation studies of collisionless relativistic reconnection have found normalized reconnection rates of~$\eta_{\rm rec} \equiv v_{\rm in}/v_{\rm out} \simeq 0.1$ \citep{Liu2015,Liu2017,Liu2020,Werner2018} and efficient particle acceleration to high energies \citep{Zenitani2001,Jaroschek2004,Zenitani2007,Zenitani2008,Cerutti2013,Cerutti2014,Sironi2014,Melzani2014,Guo2014,Guo2015,Nalewajko2015,Sironi2016,Werner2016,Werner2017,Werner2018,Schoeffler2019,Mehlhaff2020,Hakobyan2021}. In particular, several studies conducted over the last two decades have shown that magnetic reconnection in the magnetically-dominated ($\sigma_h \gg 1$) regime robustly produces power-law energy distributions of energetic particles~$f\propto \gamma^{-p}$ \citep[e.g.,][]{Zenitani2001,Zenitani2007,Zenitani2008,Jaroschek2004,Guo2014,Guo2015,Guo2019,Sironi2014,Werner2016} with~$p$ decreasing with~$\sigma_h$ and approaching~$p \sim 1$ in the ultrarelativistic limit~$\sigma_h \rightarrow \infty$. 

Several nonthermal particle acceleration mechanisms have been studied theoretically in the context of magnetic reconnection, such as the ``direct" acceleration by the parallel electric field with a finite guide magnetic field (i.e., a finite non-reversing, out-of-plane component of the magnetic field) near X-points \citep{Larrabee2003,Zenitani2005,Zenitani2008,Cerutti2013,Cerutti2014,Ball2019}, Speiser orbits in the case of zero guide field \citep{Speiser1965,Hoshino2001,Zenitani2001,Uzdensky2011,Cerutti2012,Cerutti2013,Cerutti2014,Nalewajko2015,Uzdensky2022}, Fermi acceleration \citep{Fermi1949,Drake2006,Giannios2010,Guo2014,Guo2015,Dahlin2014,Zhang2021a}, and parallel electric field acceleration in the exhaust region \citep{Egedal2013,Zhang2019}. \citet{Guo2014,Guo2015,Guo2019,Kilian2020} suggest that the acceleration mechanism primarily responsible for the formation of the power-law distributions is a Fermi mechanism, but this is still under debate. Furthermore, the exact conditions responsible for the shape of the power-law distribution require further investigation. Recently, the guide-field dependence of the power-law spectra was investigated in both two and three dimensions (2D and 3D) for both a high magnetization~$\sigma_h \gg 1$ \citep{Werner2017} and a moderate magnetization~$\sigma_h = 1$ \citep{Werner2021}, and it was found that the nonthermal particle spectra steepen (i.e., the spectral index~$p$ increases) as the guide field strengthens. Previous studies of particle acceleration have included 3D simulations, but due to effects inherent to 3D \citep{Dahlin2017,Zhang2021a} and the immense computational resources required to run 3D PIC simulations, less is understood about the mechanisms responsible for particle acceleration \citep{Jaroschek2004,Zenitani2007,Zenitani2008,Sironi2014,Guo2014,Guo2015,Guo2021,Werner2017,Comisso2019,Werner2021,Zhang2021b}.

In studying particle acceleration due to relativistic reconnection, it is important to understand two acceleration stages. The first is the injection stage ($\gamma \lesssim \gamma_{\rm inj}$), i.e., particle energization from the upstream thermal energy to the lower energy boundary of the power-law distribution. The second is the main particle energization stage: high-energy acceleration ($\gamma \gtrsim \gamma_{\rm inj}$), in which nonthermal power-law distributions and high-energy cutoffs are formed. As we discussed above, several different mechanisms have been proposed, and there is currently no consensus about which mechanism controls these processes. Additionally, it is unclear how each stage of acceleration depends on various parameters, although this has been discussed in nonrelativistic and transrelativistic reconnection studies \citep{Dahlin2014,Dahlin2017,Li2018a,Li2018b,Li2019a,Ball2019,Zhang2021a,Kilian2020}.

Let us note several quantities potentially deducible from both observation and simulations of relativistic reconnection. First, the power-law index~$p$---the parameter that has received the most attention in the literature so far---can be inferred from observational data (i.e., radiation spectra, e.g., \citealt{Kumar2015,Abdo2011,Tavani2011,Atoyan1999,Clausen-Brown2012,Mochol2015}) and obtained from numerical simulations (e.g., \citealt{Sironi2014,Guo2014,Werner2016}). It is often~$p < 4$ and can be very hard (e.g., $p \sim 1$). Second, the high-energy cutoff~$\gamma_c$ of the nonthermal particle distribution has important implications for their high-energy radiation, especially in X-ray and gamma-ray bands \citep[e.g.,][]{Werner2016,Zhang2021c,Zhang2022}. Additionally, the growth rate~$r$ of the high-energy cutoff~$\gamma_c(t) \propto t^r$ helps us narrow down what mechanisms are responsible for accelerating the most energetic particles \citep[e.g.,][]{Petropoulou2018,Hakobyan2021,Zhang2021a}. It may be observationally inferred from rapid variability at high energies. Third, calculating the acceleration efficiency~$\eta$ is of great interest as many observations find that the energies of a significant fraction of particles greatly exceed the energy of the spectral peak, suggesting that very efficient particle acceleration is present. For this endeavor, it is essential to determine the injection energy~$\gamma_{\rm inj}$.

A crucial task is determining how the above potentially observable quantities (power-law index, high-energy cutoff, acceleration efficiency, etc.) depend on system parameters such as guide-field strength and upstream magnetization. Additionally, the scaling of these variables with domain size (spatial and temporal) is essential for understanding astrophysical systems, as it determines if we can extrapolate the simulation results to large astrophysical scales. Developing a strong connection between these parameters can reveal what plasma conditions lead to particular features of nonthermal particle spectra. Elucidating these connections can help us assess the role of particle acceleration driven by relativistic reconnection in, e.g., violently flaring accreting black hole jets and coronae and neutron star magnetospheres \citep{Cerutti2013,Cerutti2014,Cerutti2015,Sironi2015,Cerutti2016,Beloborodov2017,Werner2018,Werner2019,Ball2019,Schoeffler2019,Cerutti2020,Cerutti2020b,Kilian2020,Sironi2020,Nattila2021}. This could help answer persistent mysteries in astrophysics, such as the origin of very energetic gamma-rays \citep{Zhang2022}.

In this paper, we use fully kinetic 2D particle-in-cell simulations of relativistic magnetic reconnection in collisionless pair plasmas to investigate how guide-field strength and domain size affect particle injection, high-energy particle spectra, and acceleration efficiencies. For the injection stage, seeing as previous studies have highlighted the importance of both the parallel electric field \citep{Ball2019, Sironi2020, Sironi2022} and the perpendicular electric field \citep{Guo2019,Kilian2020}, our approach considers several mechanisms simultaneously in order to reduce bias. In particular, we attempt to distinguish and assess the relative importance of three different acceleration processes: direct acceleration by the reconnection electric field, Fermi acceleration, and pickup acceleration. A detailed understanding of the injection mechanisms and their contributions is needed to construct injection models, which are useful in the context of global or large-scale simulations, where many regions can be approximated as extended current sheets. For the post-injection, high-energy stage of acceleration, we distinguish the contributions of parallel and perpendicular electric fields. These assessments cover a range of guide-field strengths and domain sizes with a fixed ambient upstream magnetization. For weak and moderate guide fields, we show that particle injection by perpendicular electric fields is more important than that by parallel electric fields and is completely dominant in the high-energy (main) acceleration phase. On the other hand, a strong guide field can suppress acceleration processes related to perpendicular electric fields, although these acceleration processes become increasingly important for larger domains.

To our knowledge, this is the first systematic study into how guide-field strength, varied independently from weak ($b_g = 0.1$) to strong ($b_g = 1.0$), affects particle injection from relativistic reconnection. Recently, particle injection has been studied in the transrelativistic regime for a proton-electron plasma with a weak guide field \citep{Ball2019,Kilian2020}, and relativistic pair plasmas using~$b_g = 0$ and~$b_g = 1$ with a different treatment in each case \citep{Sironi2022}, commented by \citealt{Guo2022_comment}. Furthermore, direct dependence of high-energy power-law spectra on guide-field strength has recently been investigated in 2D and 3D simulations \citep{Zenitani2008,Cerutti2013,Cerutti2014,Dahlin2014,Dahlin2017,Werner2017,Ball2019,Werner2021,Guo2021} in each regime of~$\sigma_h \equiv B_0^2/4\pi h$ [sub-relativistic ($\sigma_h \ll 1$), transrelativistic regime ($\sigma_h \simeq 1$), and relativistic ($\sigma_h \gg 1$)], where the ubiquitous result is that stronger guide fields steepen power-law spectra.

Throughout this paper, we use naturalized units, i.e., $m_e = c = 1$. That is, we normalize velocities to the speed of light~$c$, momenta to~$m_e c$, and energies to the electron rest energy~$m_e c^2$.

The rest of this paper is organized as follows. Section~\ref{sec:2_mechanisms} explains the different particle injection mechanisms and how they are distinguished. Section~\ref{sec:3_sim_setup} describes the setup for the simulations. Section~\ref{sec:4_results} contains the analysis and results from each simulation. Section~\ref{sec:5_discussion} discusses astrophysical applications, comparisons with previous work, and future work. Section~\ref{sec:6_conclusions} presents our main conclusions.

\section{Mechanisms of particle injection} \label{sec:2_mechanisms}

\begin{figure*}[htp!]
    \centering
    \includegraphics[width=18cm]{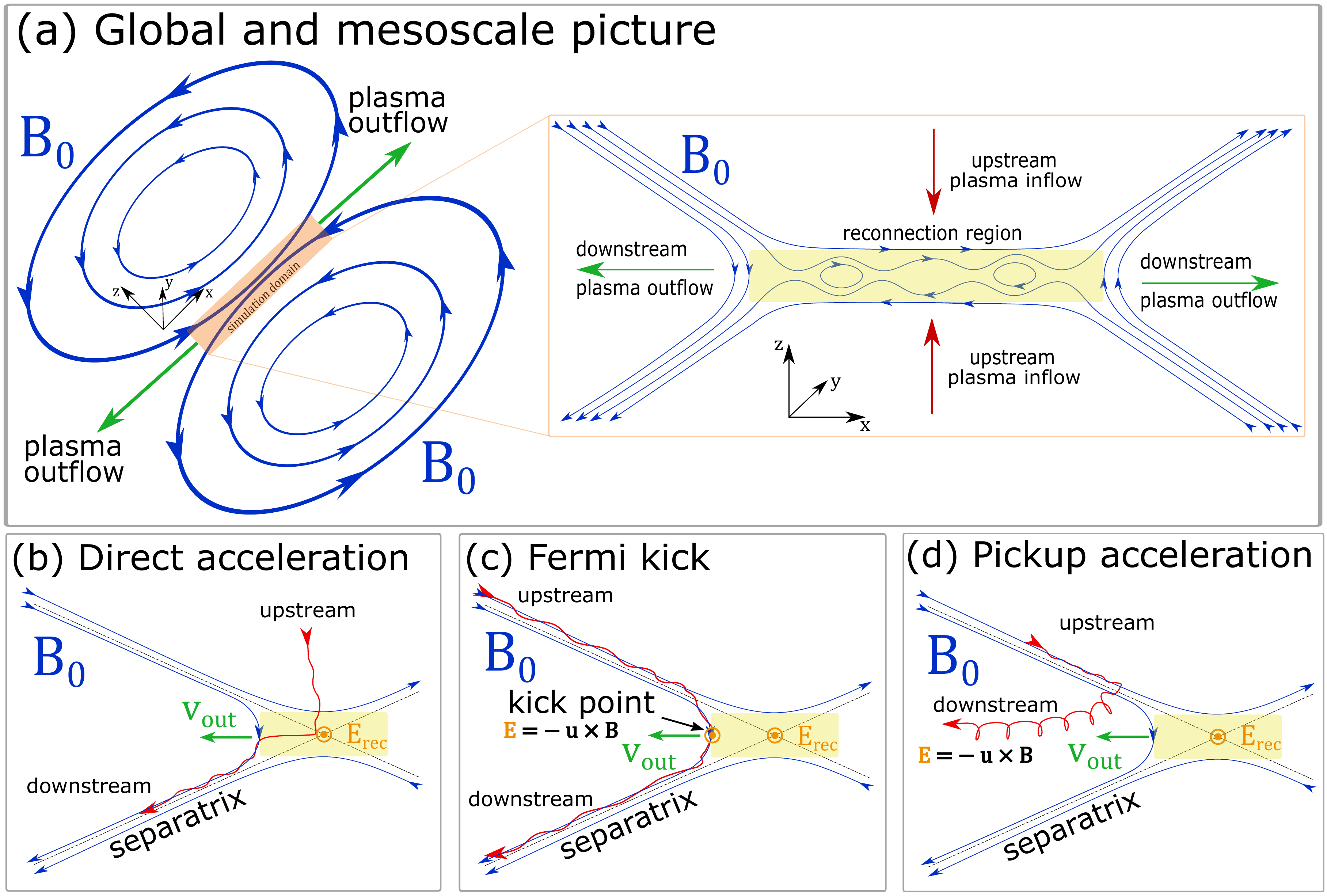}
    \caption{Sketches of global and mesoscale reconnection configurations and several particle injection mechanisms. (a) The surrounding astrophysical context of the simulation domain (highlighted in orange). (b) Injection by direct acceleration from the reconnection electric field near an X-point. (c) Injection by a Fermi ``kick." (d) Injection by the pickup process, in which~$\textbf{p}_\perp$ suddenly increases upon crossing the separatrix and subsequent entry into the downstream region. Note that the magnetic field configurations are the same for all three injection mechanisms, but the exact acceleration regions differ, even though all particles enter from the upstream and become injected in the downstream. In panels (b, c, d), $B_0$ is the reconnecting magnetic field, $E_{\rm rec}$ is the reconnection electric field, and~$v_{\rm out} \simeq v_{Ax}$ is the reconnection outflow speed, approximately equal to the in-plane Alfv\'en speed.}
    \label{fig:accel_cartoons}
\end{figure*}

Throughout this paper, the quantity~$\textbf{v}$ refers to particle 3-velocity in the simulation frame. We also define two other velocities, namely the~$\textbf{E} \times \textbf{B}$ drift velocity~$\textbf{v}_E \equiv \textbf{E} \times \textbf{B}/B^2$ and the particle velocity calculated with respect to the~$\textbf{v}_E$ frame, denoted by~$\textbf{v}'$, which is defined by:

\begin{equation} \label{equation:vprime}
    \textbf{v}' \equiv \frac{1}{1 - \textbf{v}\cdot \textbf{v}_E} \bigg( \frac{\textbf{v}}{\gamma_E} - \textbf{v}_E + \frac{\gamma_E}{1 + \gamma_E} (\textbf{v}\cdot \textbf{v}_E)\textbf{v}_E \bigg).
\end{equation}

Each velocity is associated with a Lorentz factor, e.g., 
\begin{equation} \label{equation:lorentz_factors}
    \gamma \equiv \frac{1}{\sqrt{1 - \lvert \textbf{v} \rvert^2}}, \ \ \ \gamma_E \equiv \frac{1}{\sqrt{1 - \lvert \textbf{v}_E \rvert^2}}, \ \ \ \gamma' \equiv \frac{1}{\sqrt{1 - \lvert \textbf{v}' \rvert^2}}
\end{equation}

Next, subscripts~$\parallel, \perp$ indicate velocity components relative to the local magnetic field. For example, $\textbf{v}'_\parallel$ represents the component of the particle velocity in the~$\textbf{E} \times \textbf{B}$ drift frame parallel to the local magnetic field in the~$\textbf{E} \times \textbf{B}$ drift frame.
 
We also define several momenta similar to the velocities. For example, $\textbf{p} \equiv \gamma \textbf{v}$ refers to particle momenta in the simulation frame. It can be proved that~$\textbf{p}_\parallel' = \textbf{p}_\parallel$, so we have~$\lvert \textbf{p}_\parallel \rvert = \lvert \textbf{p}_\parallel' \rvert = \gamma' \lvert \textbf{v}'_{\parallel} \rvert$. Lastly, we will use~$\lvert \textbf{p}'_\perp \rvert = \gamma' \lvert \textbf{v}'_{\perp} \rvert$ throughout the paper.

Figure~\ref{fig:accel_cartoons}a shows the broader context of the simulation box. Within the reconnection region highlighted in yellow, plasmoids form due to the tearing instability. An additional reconnection layer exists between each pair of plasmoids, which propagates this structure down in a self-similar hierarchy \citep{Shibata2001,Loureiro2007,Bhattacharjee2009,Uzdensky2010,Uzdensky2016b,Majeski2021}.

Upon entering the downstream from the cold upstream, particles may be accelerated by several different \textit{injection mechanisms}. When such particles reach an energy sufficiently greater than the background upstream thermal energy, they begin a second stage of acceleration that develops a power-law energy distribution. One such process is Fermi acceleration, driven by particle curvature drift motion along~$\textbf{E}_\perp$ \citep{Drake2006,Dahlin2014,Guo2014,Li2018b,Lemoine2019,Kilian2020,Zhang2021a}. It is convenient to define an ``injection energy" $\gamma_{\rm inj}$ that marks the beginning of the power-law distribution so that energization from $\gamma < \gamma_{\rm inj}$ to $\gamma_{\rm inj}$ is from injection mechanisms (see Appendix~\ref{sec:8_appendix_A} for more details). Recent work has identified two injection mechanisms:

\begin{enumerate}
    \item [(a)] \textbf{Direct acceleration}. Magnetic field lines undergoing reconnection induce strong electric fields around the diffusion region that accelerate particles along the reconnected magnetic field in the case of a nonzero guide field \citep[Figure~\ref{fig:accel_cartoons}b;][]{Ball2019, Kilian2020}. Direct acceleration occurs either in the initial current sheet or when two magnetic islands merge. Although not considered in this study, the case of zero guide field gives rise to Speiser-like orbits \citep{Speiser1965,Zenitani2001,Sironi2022}.
    
    \item [(b)] \textbf{Fermi kick}. 
    The relaxation of freshly-reconnected magnetic field-line tension gives rise to a universal Fermi acceleration process involving the curvature drift of particles \citep{Drake2006,Dahlin2014,Guo2014,Li2018b,Kilian2020,Zhang2021a}. The first reflection by the curved field lines injects particles \citep{Zhang2021a}. After being kicked, the particle gains momentum mainly parallel to the local magnetic field, with the magnitude of the gain depending on the Alfv\'en speed~$v_A$ (Figure~\ref{fig:accel_cartoons}c).
\end{enumerate}

In this paper, we extend this analysis of particle injection to include a third mechanism:
\begin{enumerate}
    \item [(c)] \textbf{Pickup acceleration}.\footnote{This process is somewhat analogous to the well-known heliospheric pickup process, where a neutral atom in the heliosphere becomes ionized and is suddenly picked-up by the solar wind magnetic field \citep{Mobius1985}.} When a particle suddenly enters a reconnection outflow region, it may be swept up (or ``picked up'') by the flow, achieving rapid acceleration (Figure~\ref{fig:accel_cartoons}d). This acceleration is owed to the violation of magnetic moment adiabatic invariance (henceforth, ``non-adiabatic behavior"). This process has been studied in the nonrelativistic regime \citep{Drake2009} and more recently in relativistic magnetic reconnection \citep{Sironi2020}.
\end{enumerate}

\subsection{Distinguishing the injection mechanisms} \label{ss:distinguishing_mechanisms}

In the case of a nonzero guide field, direct acceleration is well-approximated by~$\textbf{E}_\parallel$. In contrast, electrons that undergo Fermi acceleration (whether continual or a single kick) are primarily accelerated by the ``motional" electric field~$\textbf{E}_m = -\textbf{u} \times \textbf{B}$ which is induced by bulk plasma motion and is perpendicular to the local magnetic field \citep{Guo2019,Comisso2019,Kilian2020}. Generally, any perpendicular electric field that satisfies $E < B$ may support Fermi acceleration \citep{Lemoine2019}. 

This has led to a recent strategy for analysis, which is to decompose the work done by electric fields into a parallel [$W_\parallel(t) \equiv q\int_0^t \textbf{v}(t') \cdot \textbf{E}_\parallel(t') \,dt'$] and perpendicular [$W_\perp(t) \equiv q\int_0^t \textbf{v}(t') \cdot \textbf{E}_\perp(t') \,dt'$] components, where~$\textbf{E}_\parallel \equiv (\textbf{E} \cdot \textbf{B})\textbf{B}/\lvert \textbf{B} \rvert^2$ and~$\textbf{E}_\perp \equiv \textbf{E} - \textbf{E}_\parallel$ \citep{Guo2019,Comisso2019,Kilian2020}. With this decomposition, $W_\parallel$ was attributed to the reconnection electric field and~$W_\perp$ to Fermi acceleration.

However, this association subsumes the contribution of pickup acceleration to particle injection into the ``Fermi kick(s)" category because the motional electric field~$\textbf{E}_m$ also accelerates particles via the pickup process. One way to distinguish this contribution from Fermi acceleration is to compare gains in momentum that are parallel versus perpendicular to the local magnetic field; because:
\begin{enumerate}
    \item Particles accelerated by Fermi reflections gain kinetic energy via~$W_\perp$ and gain net~$\lvert \textbf{p}_\parallel \rvert$. For a single kick at low energies, the magnetic moment~$\mu' \equiv \lvert \textbf{p}_\perp' \rvert^2/2\lvert \textbf{B}' \rvert$ is conserved (i.e., the particle is adiabatic), and therefore the main momentum gain is in the parallel direction. As a direct consequence, $\lvert \textbf{p}_\parallel \rvert > \lvert \textbf{p}_\perp' \rvert$ for each Fermi kick.
    
    \item Pickup particles gain in-plane momentum $\textbf{p}_\perp \simeq \gamma_{Ax} v_{Ax}$ and have gyroradii~$\sim \gamma_{Ax} v_{Ax}/\omega_{\rm ce}$, where~$\gamma_{Ax} \equiv (1 - v_{Ax}^2)^{-1/2}$. As for their Lorentz-boosted momenta, the perpendicular velocity dominates~$\textbf{v} \simeq \textbf{v}_\perp$ and from Eq~(\ref{equation:vprime}) one finds that~$\textbf{v}' \simeq \textbf{v}'_\perp \implies \textbf{p}' \simeq \textbf{p}'_\perp \implies \lvert \textbf{p}'_\perp \rvert > \lvert \textbf{p}_\parallel \rvert$.
\end{enumerate}

We will not use this distinction to study post-injection acceleration to higher energies, as particle momenta in the turbulent reconnection layer fluctuate rapidly, likely because they are defined instantaneously and can change due to non-adiabatic motions like pitch-angle scattering. As a result, our investigation of the second energization stage ($\gamma \gtrsim \gamma_{\rm inj}$) will be limited to evaluating the contributions of~$W_\parallel$ and~$W_\perp$ (which are more stable, time-integrated quantities) to total particle energization.

In summary, the quantitative associations we propose to study particle acceleration at pre-injection energies ($\gamma \lesssim \gamma_{\rm inj}$) are:
\begin{equation} \label{association2}
\begin{gathered}
W_\parallel > W_\perp \implies \ \text{Direct acceleration by~$E_{\rm rec}$} \\
(W_\perp > W_\parallel) \ \& \ (\lvert \textbf{p}_\parallel \rvert  > \lvert \textbf{p}_\perp' \rvert) \implies \ \text{Fermi kick(s)} \\
(W_\perp > W_\parallel) \ \& \ (\lvert \textbf{p}_\parallel \rvert < \lvert \textbf{p}_\perp' \rvert) \implies \ \text{Pickup process}
\end{gathered}
\end{equation}

The procedure by which a given particle is categorized into an injection mechanism is as follows. First, we track the relevant quantities of the particle (i.e., $\gamma$, $W_\parallel$, $W_\perp$, $\lvert \textbf{p}_\parallel \rvert$, $\lvert \textbf{p}_\perp' \rvert$) until the time~$t_{\rm inj}$, defined as when $\gamma \geq \gamma_{\rm inj}$ is first satisfied (where~$\gamma_{\rm inj}$ is determined using a fitting routine for each simulation at the final time; see Appendix~\ref{sec:8_appendix_A}). At~$t = t_{\rm inj}$, the conditions defined in Eq.~(\ref{association2}) are considered, and the particle is categorized according to whichever condition it satisfies.

While the procedure attributes one injection mechanism to each injected particle, we note that multiple injection mechanisms can work together toward injecting a single particle. Furthermore, since the momenta~$\lvert \textbf{p}_\parallel \rvert$ and~$\lvert \textbf{p}_\perp' \rvert$ are defined instantaneously, some particles may be misclassified between Fermi and pickup acceleration. Therefore, we run the mechanism classification procedure over many particles to gain a statistical measure of the contribution of each injection mechanism.

\section{Simulation Setup} \label{sec:3_sim_setup}

To study particle injection and acceleration by relativistic magnetic reconnection, we perform an array of 2D, collisionless, fully kinetic simulations using the \code{VPIC} code, which solves the relativistic Vlasov-Maxwell equations \citep{Bowers2008}. In all of our simulations, we start with a force-free current sheet (CS).

The plasma density is normalized by the initial plasma density $n_0 = n_e + n_i$, which is uniform and represented by $100$ positron-electron pairs per computational grid cell. The pair mass ratio is~$m_i/m_e = 1$. Currents are normalized by~$en_0c/2$. The initial plasma is thermal and relativistically cold with a uniform temperature~$\theta \equiv T/m_ec^2 = 0.25$. The cold upstream magnetization parameter is set to~$\sigma \equiv B_0^2/4\pi n_0m_ec^2 = 50$, where~$B_0$ is the reconnecting magnetic field. Since~$\theta$ is sub-relativistic, this~$\sigma$ is close to the ``hot" upstream magnetization~$\sigma_h \equiv \sigma/\theta$. Both~$B_0$ and~$n_0$ (and hence $\sigma$) are held fixed across all simulations.

In this study, we characterize system sizes by the dimensionless measure~$\ell \equiv L/\sigma \rho_0$, where~$L$ is the system size and~$\rho_0 \equiv c/\omega_{\rm ce} = m_e c^2/e B_0$ is the nominal relativistic gyroradius and $\omega_{\rm ce}$ is the electron-cyclotron frequency in the reconnecting magnetic field without relativistic correction. It is also customary to characterize the system size by the dimensionless measure~$\ell = L/d_e$, where~$d_e \equiv c/\omega_{\rm pe}$ is the initial nonrelativistic plasma skin depth and~$\omega_{\rm pe} \equiv \sqrt{4\pi n_e e^2/m_e}$ is the plasma-electron frequency. This can be related to~$\sigma \rho_0$ via~$d_e = \sqrt{\sigma}\rho_0$. All of our simulations are run within rectangular boxes in the $x$-$z$-plane with~$x \in [0, \ell_x]$, $z \in [-\ell_z/2, \ell_z/2]$, with~$y$ as the ignored coordinate. The aspect ratio is fixed to~$\ell_x/\ell_z \equiv 2.048$. To resolve kinetic scales, the resolution is set to $\Delta x = \Delta z = d_e/4 \simeq \sigma \rho_0/28$. In the~$x$-direction, periodic boundary conditions are set for fields and particles, while in the~$z$-direction, conducting boundaries are set for fields and reflecting boundaries are set for particles. The simulation running time for most simulations is~$\tau_f = 2L_x/c$ and, when~$b_g = 1.0$, we use~$\tau_f = 4L_x/c$.\footnote{As guide fields strengthen, the in-plane component of the Alfv\'en speed becomes smaller, which significantly delays the time for developing power-law energy spectra with steady indices.} We do not use any additional filtering in our simulations.

Our simulations examine the effects of two parameters independently. The first is the ``guide-field strength'', for which we use four values, $b_g \equiv B_g/B_0 \in \{ 0.1, 0.3, 0.5, 1.0\}$, where~$B_g$ is the guide magnetic field (i.e., a uniform magnetic field pointing normal to the reconnection plane). The second is the domain size, for which we use six values, $\ell_x \in \{ 72.4, 101.8, 144.8, 203.6, 289.6, 407.3 \}$, corresponding to $\{ 2048, 2880, 4096, 5760, 8192, 11520 \}$ grid cells in the~$x$-direction.

The initial magnetic field is
\begin{equation} \label{eq:mag_field}
   \textbf{B} \equiv B_0 \tanh{(z/\lambda)} \ \hat{\textbf{x}} + B_0 \sqrt{\sech^2{(z/\lambda)} + b_g^2} \  \hat{\textbf{y}}, 
\end{equation}
where~$\lambda$ is the half-thickness of the initial CS (set to~$6\, d_e \simeq 0.85\, \sigma \rho_0$). We add a small initial perturbation to trigger magnetic reconnection, identical to \citet{Kilian2020}. \footnote{Using a similar setup to this work, \citet{Werner2021} found that, in 2D, while increasing the initial perturbation strength hastens the onset of reconnection, the subsequent energy conversion evolves almost identically to the case of zero perturbation.} While we do not exclude initial current-carrying particles, we expect their influence on the resulting particle acceleration to be negligible, as concluded by \citet{Kilian2020}.

In this paper, we wish to investigate particle spectra in the downstream\footnote{``Downstream" refers to the region between two separatrices.} region. Therefore, we isolate the downstream region by particle mixing \citep{Daughton2014}. A mixing fraction~$\mathcal{F}_e \equiv (n_e^{\rm bot} - n_e^{\rm top}) / (n_e^{\rm bot} + n_e^{\rm top})$ that satisfies~$\lvert \mathcal{F}_e \rvert \leq 99\%$ defines the downstream region and~$\lvert \mathcal{F}_e \rvert > 99\%$ defines the upstream.\footnote{Here, $n_e^{\rm bot}$ and~$n_e^{\rm top}$ are the number densities of electrons that start at~$z < 0$ and~$z > 0$, respectively.}

In analyses pertinent to particle injection (see Section~\ref{ss:inj_results}), we uniformly select~$\sim 400,000$ particles at the beginning of each simulation and track relevant physical quantities associated with them at each time step, such as positions, velocities, and electric and magnetic fields. From this information, we statistically analyze the acceleration mechanisms of particles \citep{Guo2016,Guo2019,Li2019a,Li2019b}.

\begin{figure*}[htp!]
    \centering
    \includegraphics[width = 18cm]{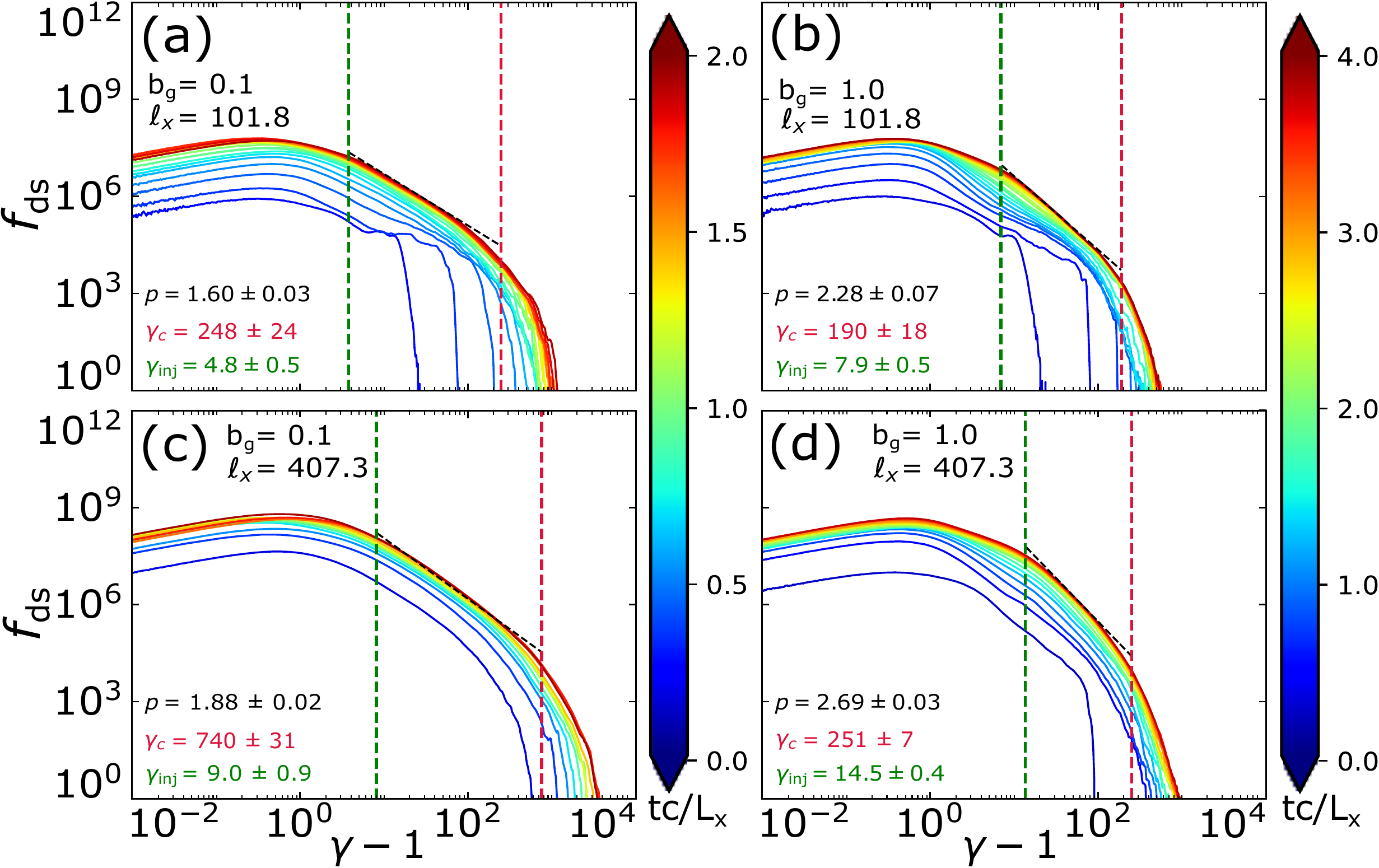}
\caption{Time evolution of the downstream electron spectra in four different simulations: (a)~$\ell_x = 101.8$, $b_g = 0.1$, (b)~$\ell_x = 101.8$, $b_g = 1.0$, (c)~$\ell_x = 407.3$, $b_g = 0.1$, and (d)~$\ell_x = 407.3$, $b_g = 1.0$. Stronger guide fields slow down the formation of the power laws, as indicated by the number of light crossing times on each color bar. The green and red dashed vertical lines indicate~$\gamma_{\rm inj}$ and~$\gamma_c$ at the end of each simulation.}
\label{spectra}
\end{figure*}

\vspace{1cm}

\section{Analysis and Results} \label{sec:4_results}

The results will be divided into four parts. Subsection~\ref{ss:spectra} discusses particle spectra and its features; Subsection~\ref{ss:accel_efficiency} is about acceleration efficiencies; we discuss particle injection in Subsection~\ref{ss:inj_results} and post-injection acceleration in Subsection~\ref{ss:postinj_results}.

\subsection{Particle spectra} \label{ss:spectra}

\begin{figure*}[htp!]
    \centering
    \includegraphics[width = 18cm]{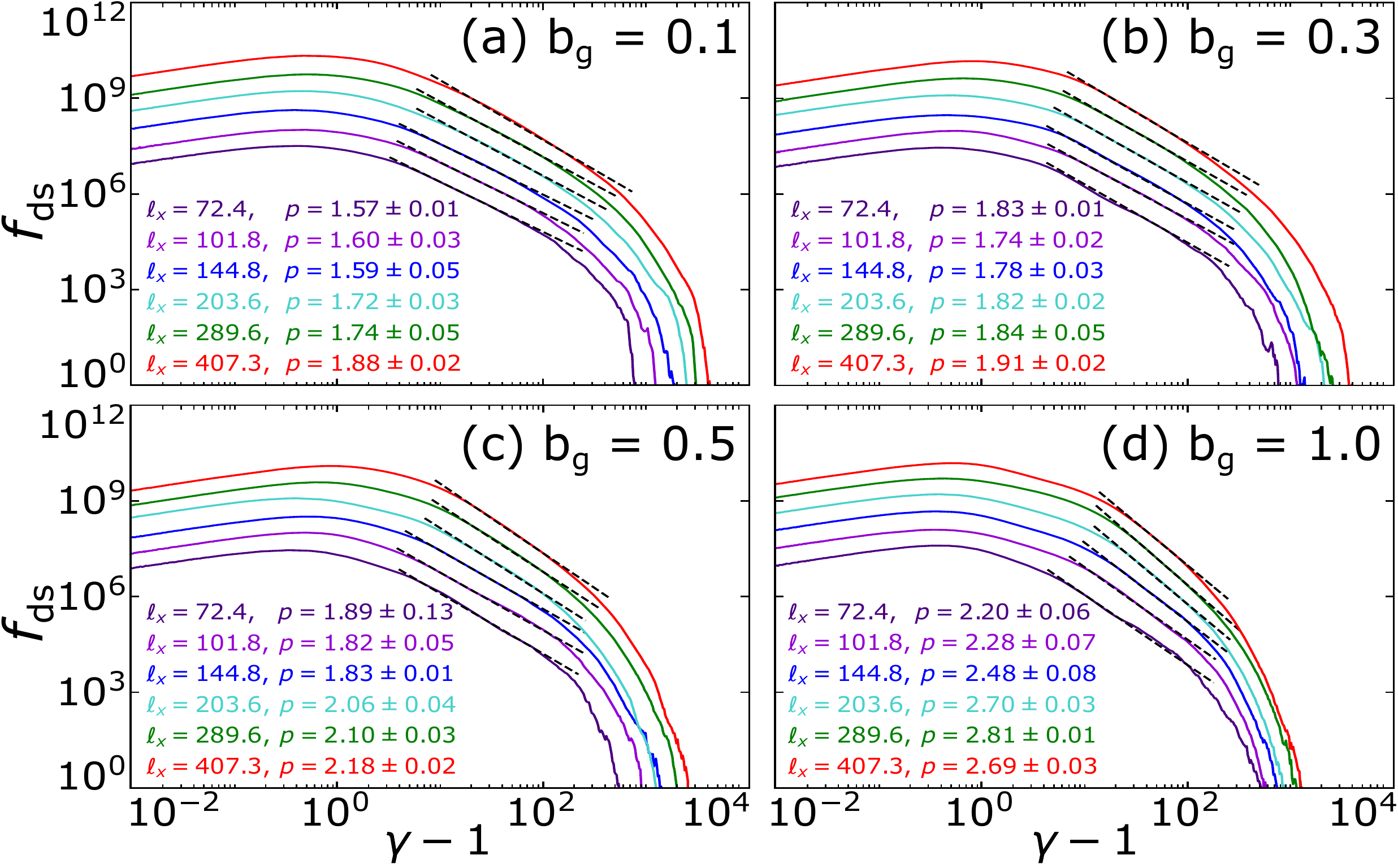}
    \caption{Downstream electron spectra at final times ($t = \tau_f$) over the entire parameter scan. The dashed lines shown indicate the fitted power-law segments, with their endpoints constrained by~$\gamma_{\rm inj}$ and~$\gamma_c$.}
    \label{fig:spect_t=tf}
\end{figure*}

\begin{figure*}[htp!]
    \centering
    \includegraphics[width = 18cm]{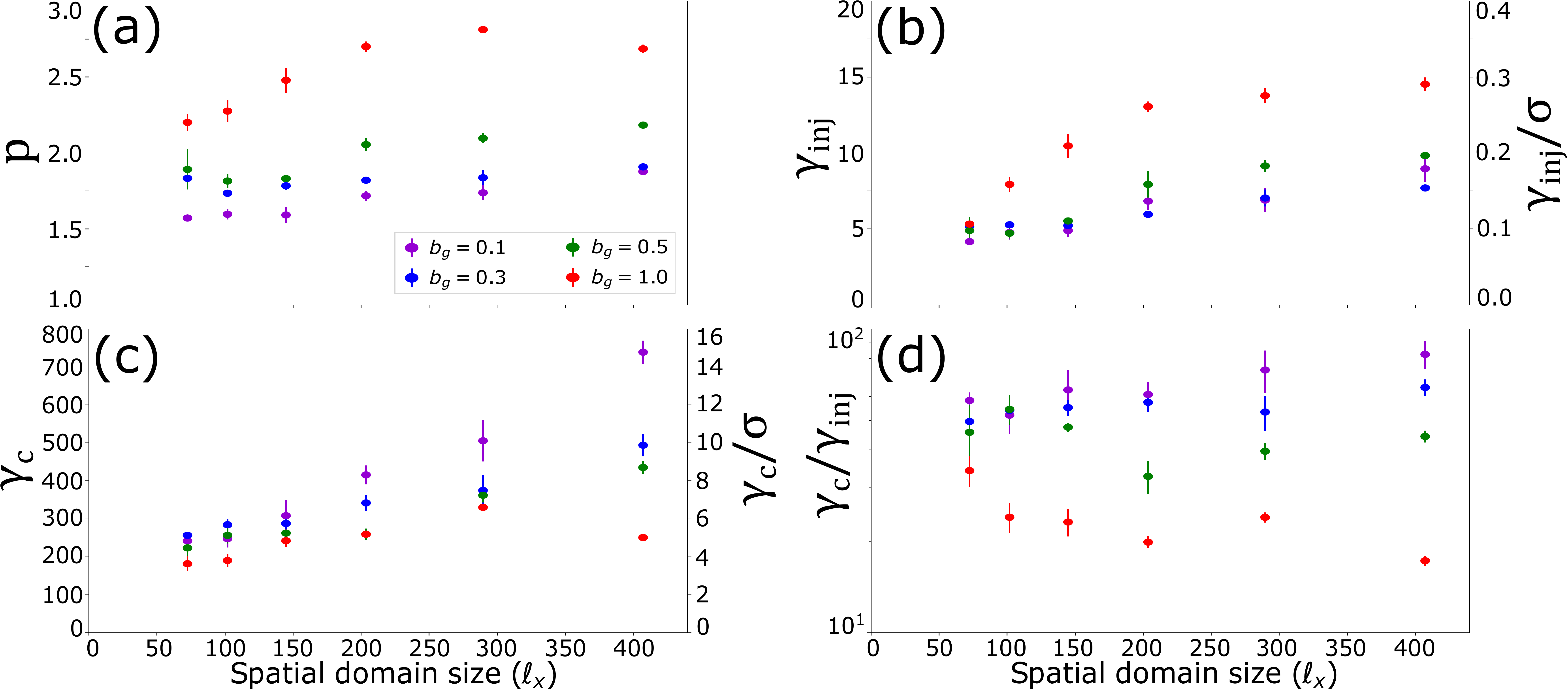}
    \caption{Several key nonthermal spectrum parameters obtained from the power-law fitting procedure (Appendix~\ref{sec:8_appendix_A}) at the final times plotted against domain size for different guide-field strengths: (a) the power-law index~$p$, (b) the injection energy~$\gamma_{\rm inj}$, (c) the high-energy cutoff~$\gamma_c$, (d) the power-law extent~$\gamma_c/\gamma_{\rm inj}$.}
    \label{spectra_results}
\end{figure*}

The electron spectrum of the downstream region can be separated into two components. The first component is a thermal component, which can be approximated by a Maxwellian-like distribution. The second component is the high-energy nonthermal component that is well-approximated by a power law with a high-energy cutoff~$f_{\rm nt}(\gamma) \propto \gamma^{-p}\, e^{-\gamma/\gamma_c}$ for~$\gamma \gtrsim \gamma_{\rm inj}$, where~$\gamma_c$ is the high-energy cutoff. Appendix~\ref{sec:8_appendix_A} details how~$\gamma_{\rm inj}$ and~$\gamma_c$ are calculated. Since our simulations are 2D, nearly all high-energy ($\gamma > \gamma_{\rm inj}$) particles are located in the downstream region (i.e., in reconnection exhausts and plasmoids).

Understanding the time evolution of particle spectra has strong implications for observations; in particular, whether various spectral features depend directly on the domain size~$\ell_x$. Figure~\ref{spectra} shows the time evolution of downstream particle spectra, for two cases of guide-field strength ($b_g = 0.1, 1.0$) and domain size ($\ell_x = 101.8, 407.3$). After the reconnection onsets, the nonthermal power-law distribution is established rapidly (i.e., independent of domain size); within $\omega_{\rm pe}t \sim 500$. Afterwards, the corresponding power-law index~$p$ stabilizes (within~$\sim 0.4\, L_x/c$ for~$b_g = 0.1$ and~$\sim 1.2\, L_x/c$ for~$b_g = 1.0$ provided that a sufficiently large domain is used), while the high-energy cutoff~$\gamma_c$ grows as the simulation proceeds.

Figure~\ref{fig:spect_t=tf} shows every particle spectrum at the final times~$t = \tau_f$ over the parameter scan. The dashed lines represent the power-law fits that begin at~$\gamma = \gamma_{\rm inj}$ and end at~$\gamma = \gamma_c$. The values and uncertainty estimates for~$p$, $\gamma_{\rm inj}$, $\gamma_c$, and~$\gamma_c/\gamma_{\rm inj}$ are shown in Figure~\ref{spectra_results}.

We find that the final power-law index increases in a seemingly convergent fashion as~$\ell_x$ increases (Figure~\ref{spectra_results}a). This is broadly consistent with previous studies in the relativistic regime \citep{Werner2016,Werner2018}, where~$p(t)$ for a given simulation was found to converge with time, as well as other studies where~$p(\tau_f)$ was found to converge as~$\ell_x$ increases \citep{Guo2014,Guo2015,Werner2017,Werner2018}. This is also consistent with~\citealt{Ball2018} and~\citealt{Werner2021}, who found a similar trend in the transrelativistic regime ($\sigma_h = 1$). However, due to the much greater upstream magnetization of~$\sigma_h \simeq \sigma = 50$, the simulations presented here produce considerably harder spectra, with~$p$ ranging from~$\sim 1.5$ to~$\sim 3.0$  depending on~$b_g$ (e.g., in expected contrast to~$p$ ranging from~$\sim 4.0$ to~$\sim 11.0$ in $\sigma_h = 1$ simulations by \citealt{Werner2021}).

\begin{figure}[tp!]
    \centering
    \includegraphics[width = 8.5cm]{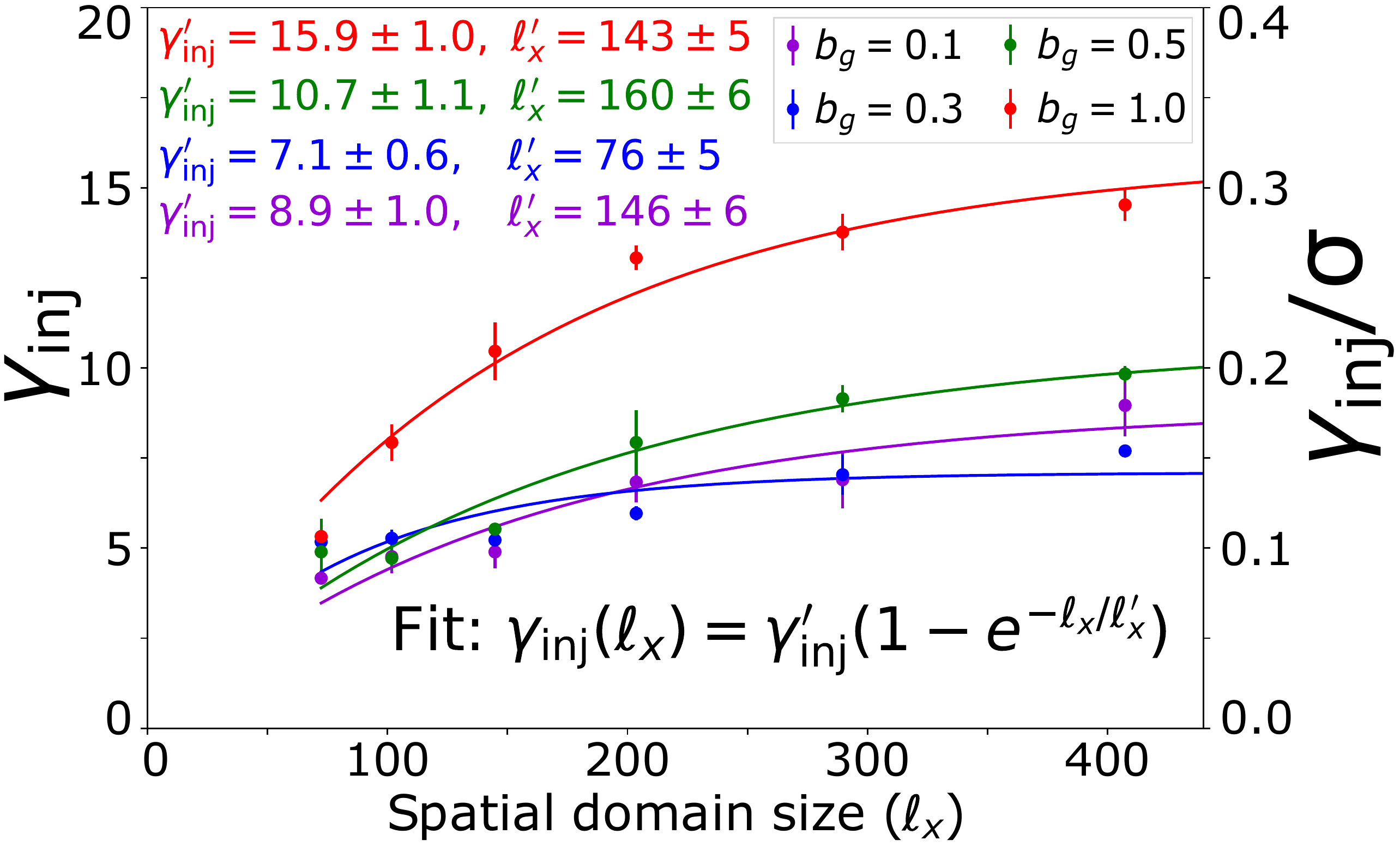}
    \caption{The $\ell_x$-dependence of the injection energy~$\gamma_{\rm inj}$ fitted with the curve~$\gamma_{\rm inj}(\ell_x) = \gamma_{\rm inj}'(1 - e^{-\ell_x/\ell_x'})$, where~$\gamma_{\rm inj}'$ and~$\ell_x'$ are $b_g$-dependent fitting parameters.}
    \label{fig:gamma_inj_fitted}
\end{figure}

Now let us move on to the $\ell_x$-dependence of the injection energy $\gamma_{\rm inj}$ (Figure~\ref{spectra_results}b). Given the apparent convergent behavior for~$\gamma_{\rm inj}$, we decided to perform a fit~$\gamma_{\rm inj}(\ell_x) = \gamma_{\rm inj}'(1 - e^{-\ell_x/\ell_x'})$ to obtain estimates for the limiting values, as shown in Figure~\ref{fig:gamma_inj_fitted}. Under this fit, $\gamma_{\rm inj}'$ represents the converging value for the injection energy as~$\ell_x$ increases, and~$\ell_x'$ measures which domain sizes are necessary to achieve injection energies close to that limiting value. We have reported~$\gamma_{\rm inj}'$ and~$\ell_x'$ for each $b_g$ in Figure~\ref{fig:gamma_inj_fitted}. We find that, in the limit of increasing~$\ell_x$, $\gamma_{\rm inj}/\sigma$ approaches $\sim 0.15$ for a weak guide field of~$b_g = 0.1$ and~$\sim 0.30$ for a strong guide field of~$b_g = 1.0$. For each value of guide-field strength, except for one outlier ($b_g = 0.3$), we find that~$\ell_x' \sim 150$. See Table~\ref{tab:gamma_inj} for more details. 

Moving on to the high-energy cutoff~$\gamma_c$, we note that fits of~$\gamma_c(t)$ or~$\gamma_c(\ell_x)$ are not reported in this study, as most of our simulations did not have a sufficient running time for the time-based growth rate~$r$ of~$\gamma_c \sim t^r$ to stabilize (as reported by \citealt{Petropoulou2018}). Nevertheless, some important qualitative observations can be made about~$\gamma_c$ variation across the parameter scan (Figure~\ref{spectra_results}c). First, it appears that~$\gamma_c(\ell_x)$ at the final time grows with increasing~$\ell_x$ for~$b_g = 0.1, 0.3, 0.5$. This suggests that~$\gamma_c$ is only limited by~$\ell_x$, which is in broad agreement with \citet{Zhang2021a} but in some disagreement with \citet{Werner2016}, who instead suggested that~$\gamma_c$ is limited to a finite multiple of~$\sigma$, with a weakening $\ell_x$-dependence in the asymptotic limit of increasing $\ell_x$. Therefore, an interesting and relevant question is how the system-size growth rate of~$\gamma_c$ depends on~$b_g$. From inspection of the~$b_g = 0.1$ case (i.e., the purple marks in Figure~\ref{spectra_results}c), it appears that the growth rate of~$\gamma_c$ is roughly consistent with linear growth. As the guide field strengthens, the growth rate of~$\gamma_c$ declines. In the~$b_g = 1.0$ case (using~$\tau_f = 4L_x/c$), $\gamma_c$ saturates with increasing~$\ell_x$ to~$\gamma_c \simeq 250 = 5\sigma$. When using~$\tau_f = 2L_x/c$, we also find saturation (to~$\gamma_c \sim 200$). See Table~\ref{tab:gamma_c} for more details. 

The power-law extents (dynamic ranges)~$\gamma_c/\gamma_{\rm inj}$ all fall within 1-2 decades. Since the injection energy seems to converge as~$\ell_x$ increases (for~$b_g = 0.1, 0.3, 0.5, 1.0$), while the cutoff energy keeps growing, for large~$\ell_x$ we expect the power-law extents to grow in accordance with~$\gamma_c$.

Finally, we discuss the guide-field dependence of the nonthermal power-law parameters. As the guide field strengthens, we find the power-law indices to increase, consistent with~\citet{Werner2017,Werner2021}, with the final asymptotic (as $\ell_x \to \infty$) values of~$p$ approaching $\sim 1.90$, $\sim 1.90$, $\sim 2.20$, and~$\sim 2.80$ for $b_g = 0.1, 0.3, 0.5, 1.0$, respectively. When varying~$b_g$ from~$0.1$ to~$0.3$, the power-law index~$p$ does not change much, in agreement with previous studies \citep{Werner2017}. Panels~(b, c) of Figure~\ref{spectra_results} suggest that~$\gamma_{\rm inj}$ and~$\gamma_c$ have opposite trends with respect to changing guide-field strength~$b_g$: $\gamma_{\rm inj}$ grows with stronger~$b_g$, while~$\gamma_c$ declines with stronger~$b_g$. As a result, the dynamic ranges of the power-law segments are shortened (i.e., their extents~$\gamma_c/\gamma_{\rm inj}$ decline) from both sides as the guide field strengthens (panels b, c, d of Figure~\ref{spectra_results}). See Appendix~\ref{sec:9_appendix_B} for more details.

\subsection{Acceleration efficiency} \label{ss:accel_efficiency}

\begin{figure*}[htp!]
    \centering
    \includegraphics[width = 18cm]{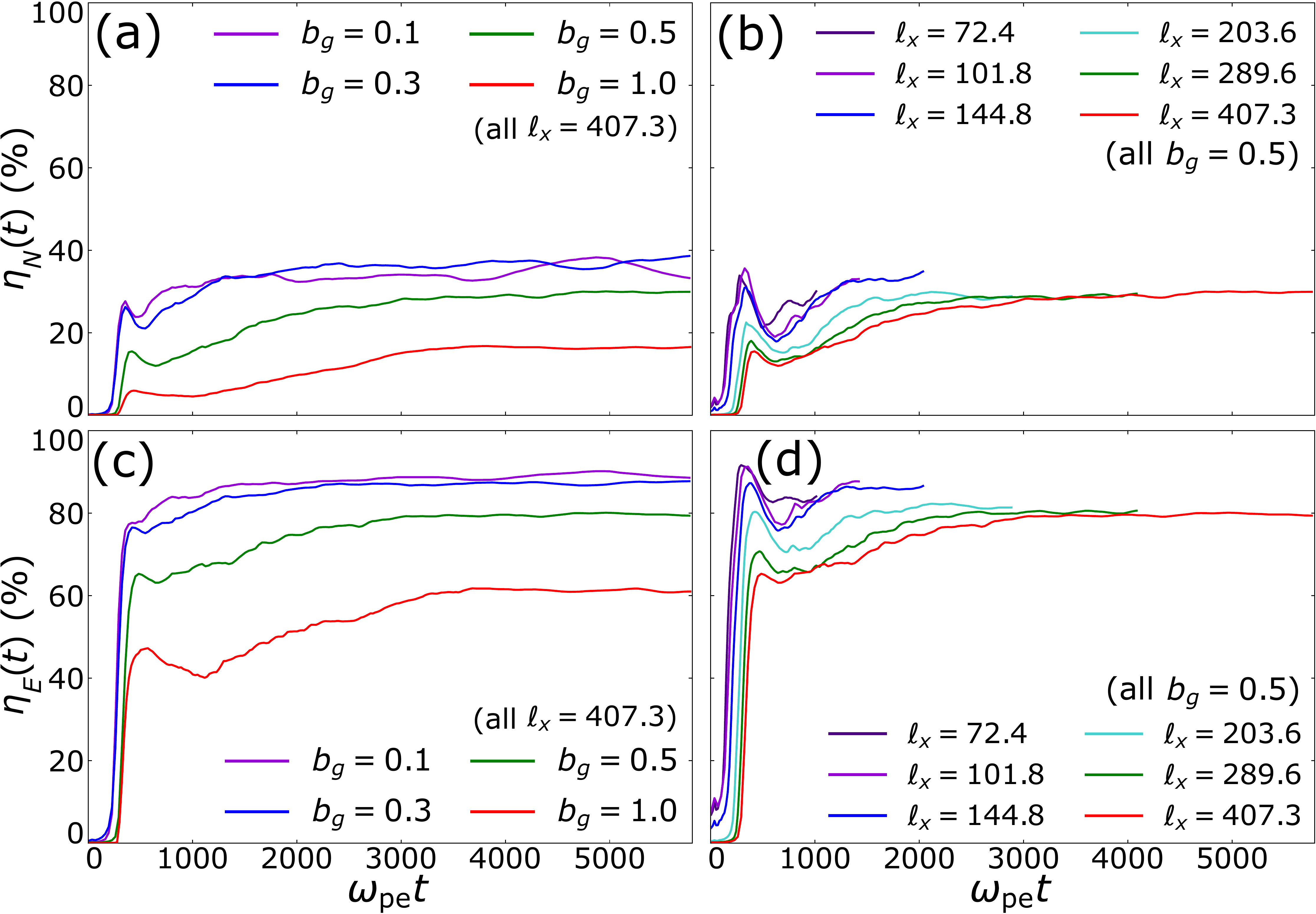}
    \caption{Evolution of acceleration efficiencies. Panels (a, b) show the number efficiency~$\eta_N(t) \equiv N_{\rm inj}/N_{\rm ds}$ [see Eq.~(\ref{equation:n_inj})] over the~$b_g$ scan with~$\ell_x = 407.3$ fixed (panel~a) and over the~$\ell_x$ scan with~$b_g = 0.5$ fixed (panel~b). Panels (c, d) show the evolution of the energy efficiency~$\eta_E(t) \equiv E_{\rm inj}/E_{\rm ds}$ [Eq.~(\ref{equation:e_inj})] over the~$b_g$ scan with~$\ell_x = 407.3$ fixed (panel~c) and over the~$\ell_x$ scan with~$b_g = 0.5$ fixed (panel~d). Note that uncertainties in~$\eta_N(t)$ and~$\eta_E(t)$ (propagated from~$\gamma_{\rm inj}$ uncertainties) are not shown but are~$\lesssim 5\%$.}
    \label{fig:efficiencies}
\end{figure*}

\begin{figure*}[bp!]
    \centering
    \includegraphics[width = 18cm]{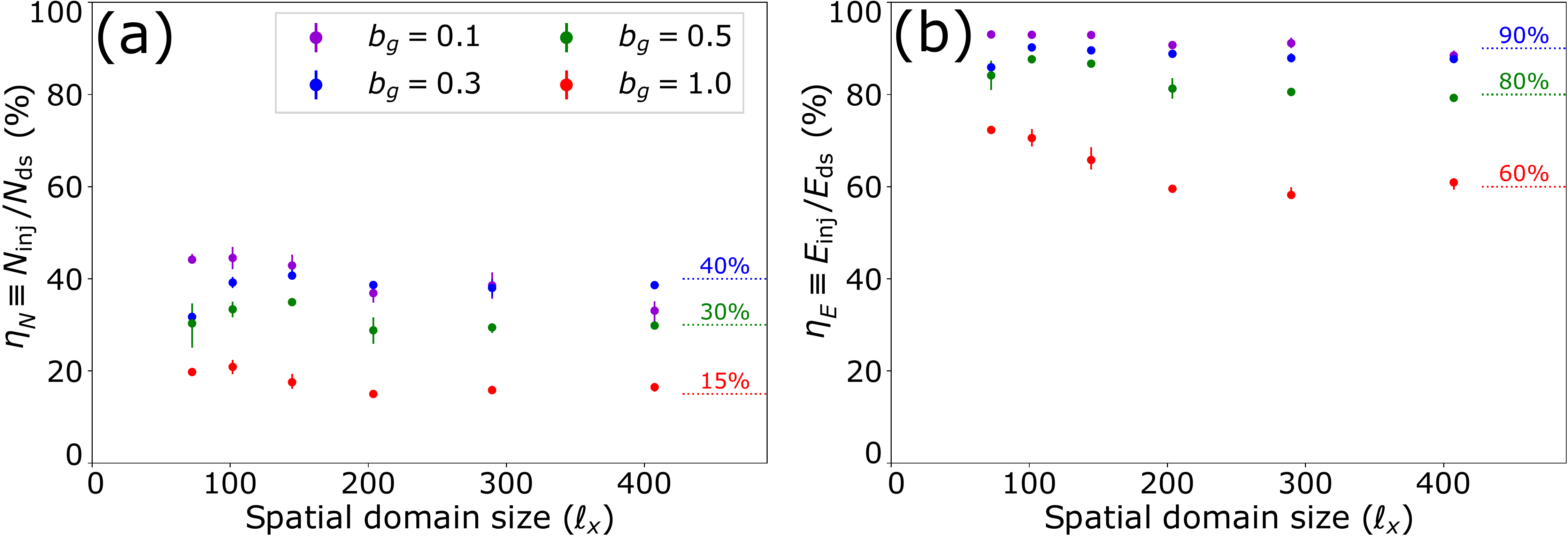}
    \caption{Acceleration efficiencies at~$t = \tau_f$ across the parameter scan. (a) The number efficiency~$\eta_N \equiv N_{\rm inj}/N_{\rm ds}$ [Eq.~(\ref{equation:n_inj})]. (b) The energy efficiency~$\eta_E \equiv E_{\rm inj}/E_{\rm ds}$ [Eq.~(\ref{equation:e_inj})]. Uncertainties are propagated from~$\gamma_{\rm inj}$ (Figure~\ref{spectra_results}b).}
    \label{fig:accel_efficiencies_tf}
\end{figure*}

Previous studies have determined relativistic magnetic reconnection to be an efficient source of nonthermal particles (e.g., \citealt{Zenitani2001,Zenitani2005,Zenitani2007,Zenitani2008,Jaroschek2004,Sironi2014,Guo2014,Werner2016}), but the acceleration efficiency has only recently been systematically studied with definite notions \citep{Hoshino2022}. We define two notions of acceleration efficiency in a fashion similar to that in \citet{Hoshino2022}. The first we call the ``number efficiency," which is the fraction of downstream electrons that have been injected, i.e., $\eta_N \equiv N_{\rm inj}/N_{\rm ds}$, where
\begin{equation} \label{equation:n_inj}
    N_{\rm ds} \equiv \int_1^\infty f_{\rm ds}(\gamma) \,d\gamma, \ N_{\rm inj} \equiv \int_{\gamma_{\rm inj}}^\infty f_{\rm ds}(\gamma) \,d\gamma.
\end{equation}
The second we call the ``energy efficiency," which is the fraction of downstream particle energy contained by injected particles, i.e., $\eta_E \equiv E_{\rm inj}/E_{\rm ds}$, where
\begin{equation} \label{equation:e_inj}
    E_{\rm ds} \equiv \int_1^\infty \gamma f_{\rm ds}(\gamma) \,d\gamma, \ E_{\rm inj} \equiv \int_{\gamma_{\rm inj}}^\infty \gamma f_{\rm ds}(\gamma) \,d\gamma.
\end{equation}
We note that~$\eta_N$ and~$\eta_E$ have their associated uncertainties, mostly owing to the propagation of the uncertainties in~$\gamma_{\rm inj}$ (see Appendix~\ref{sec:8_appendix_A} and Figure~\ref{spectra_results}b for details). 

The number efficiency~$\eta_N$ and the energy efficiency~$\eta_E$ are quantities of particular theoretical interest. First, the time evolution of~$\eta_N$ and its limiting value as~$\ell_x$ increases are useful for calculating the overall contributions of injection mechanisms. Correspondingly, in Section~\ref{ss:inj_results} we study particle injection by evaluating the \textit{number} of injected particles (as a function of time, at final times, for each injection mechanism delineated in Eq.~(\ref{association2})). A complementary quantity is the time evolution of~$\eta_E$, as its limiting value as~$\ell_x$ increases has implications for the efficiency of post-injection acceleration mechanisms ($\gamma \gtrsim \gamma_{\rm inj}$) responsible for power-law formation. Correspondingly, in Section~\ref{ss:postinj_results} we evaluate energization from parallel and perpendicular electric fields for~$\gamma > \gamma_{\rm inj}$.

Figure~\ref{fig:efficiencies} shows the time evolution of~$\eta_N$ and~$\eta_E$ for various guide-field strengths (panels a and~c) and domain sizes (panels b and~d). In all cases, we see that both~$\eta_N$ and $\eta_E$ initially rise very rapidly to a first peak, after which they experience a relatively quick moderate drop (for small domains), followed by a gradual rise to a late-time asymptotic saturation value. Panels~(a, c) show that both the initial peaks and the final saturation values of these efficiencies decline with an increased guide field, and the timescales of their late-time rise become longer. Panels~(b, d) show that the collective timescale of the initial rise and peak consistently appears to  be~$\omega_{\rm pe}t \sim 500$, independent of the system size (at a fixed $b_g=0.5$); this suggests that these features follow from the initial reconnection phase. On the other hand, when considering the late-time stage of efficiency saturation to the final asymptotic value, for a sufficiently large domain\footnote{We note that for domains that are too small (e.g., $\ell_x = 72.4$ when~$b_g = 0.5$), there is not enough time for convergence to be established.}, both~$\eta_N$ and~$\eta_E$ converge on the macroscopic dynamical timescale ($\sim L_x/c$), which is scalable to larger domains. This puts forward a picture that the timescale required to achieve the limiting efficiency scales with the length of the current sheet. Physically, the timescales of the initial features (the rapid rise and peak) correspond to those for power-law \textit{formation}, whereas the timescales of saturation correspond to those for power-law \textit{stabilization}\footnote{The timescale of power-law ``stabilization" is the time required for the power-law index~$p(t)$ to settle on a time-independent value close to~$p(\tau_f)$.} (see Figure~\ref{spectra}). This suggests that the timescale of power-law formation is transient (i.e., independent of~$\ell_x$), whereas the timescale of power-law stabilization scales with the length of the current sheet~$\ell_x$ (and also depends on the guide-field strength~$b_g$).

Finally, Figure~\ref{fig:accel_efficiencies_tf} shows the values of~$\eta_N$ and~$\eta_E$ at final times~$\tau_f$ for the complete parameter scan of~$b_g$ and~$\ell_x$. By inspection, it appears that~$\eta_N(\tau_f)$ and~$\eta_E(\tau_f)$ rapidly achieve convergence with increasing~$\ell_x$. When varying~$b_g$ from~$0.1$ to~$1.0$, we find that~$\eta_N(\tau_f)$ declines significantly (from~$\sim 40\%$ to ~$\sim 15\%$), and~$\eta_E$ declines significantly as well, from~$\sim 90\%$ to ~$\sim 60\%$. These trends highlight the importance of guide-field strength in suppressing both number and energy efficiency.

\subsection{Particle injection} \label{ss:inj_results}

\begin{figure*}[htp!]
    \centering
    \includegraphics[width = 1\textwidth]{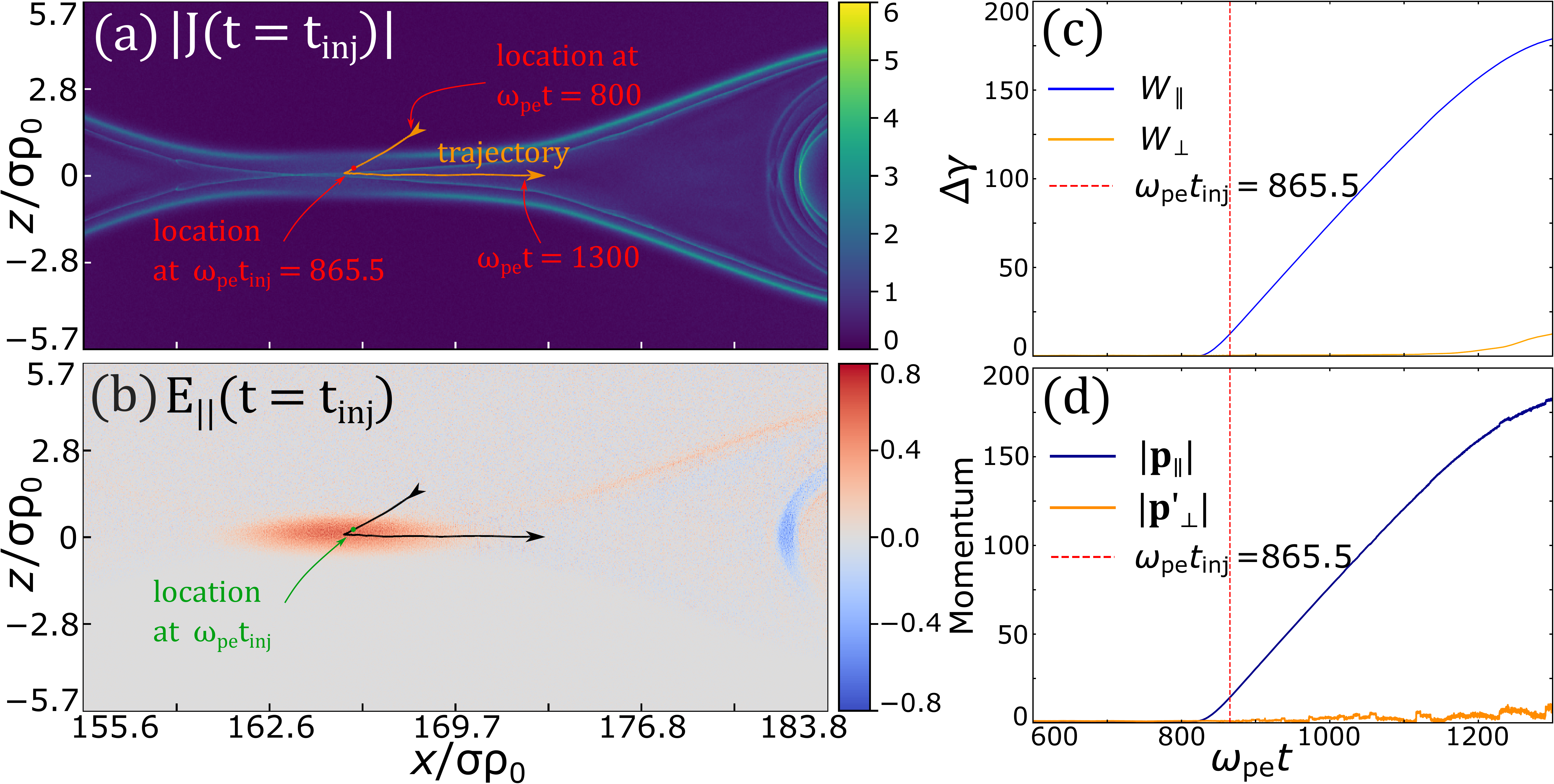}
    \caption{Example of particle injection by the reconnection electric field. Here, $\ell_x = 289.6$ and~$b_g = 1.0$. (a): Current density magnitude~$|J|$ color map at~$t = t_{\rm inj}$ with the particle trajectory shown in orange. (b): Parallel electric field~$E_\parallel$ color map at~$t = t_{\rm inj}$ with the particle trajectory shown in black. (c) Work done on the particle by parallel and perpendicular electric fields as a function of time. (d) Components of the particle's momentum as a function of time.} 
    \label{fig:direct_accel_example}
\end{figure*}

In this subsection, we will show the relative contributions of direct acceleration, Fermi reflection, and pickup process (discussed in Section~\ref{sec:2_mechanisms}) to particle injection. In particular, we study how each contribution varies with guide-field strength and system size. We categorize each injected particle into one of these mechanisms (according to Eq.~(\ref{association2})). We evaluate the relevant quantities at~$t = t_{\rm inj}$, i.e., when the particle reaches the injection energy~$\gamma_{\rm inj}$ for the first time ($\gamma_{\rm inj}$ values have been discussed in Section~\ref{ss:spectra}).

\subsubsection{Particle trajectories} 

\begin{figure*}[htp!]
    \centering
    \includegraphics[width = 17.7cm]{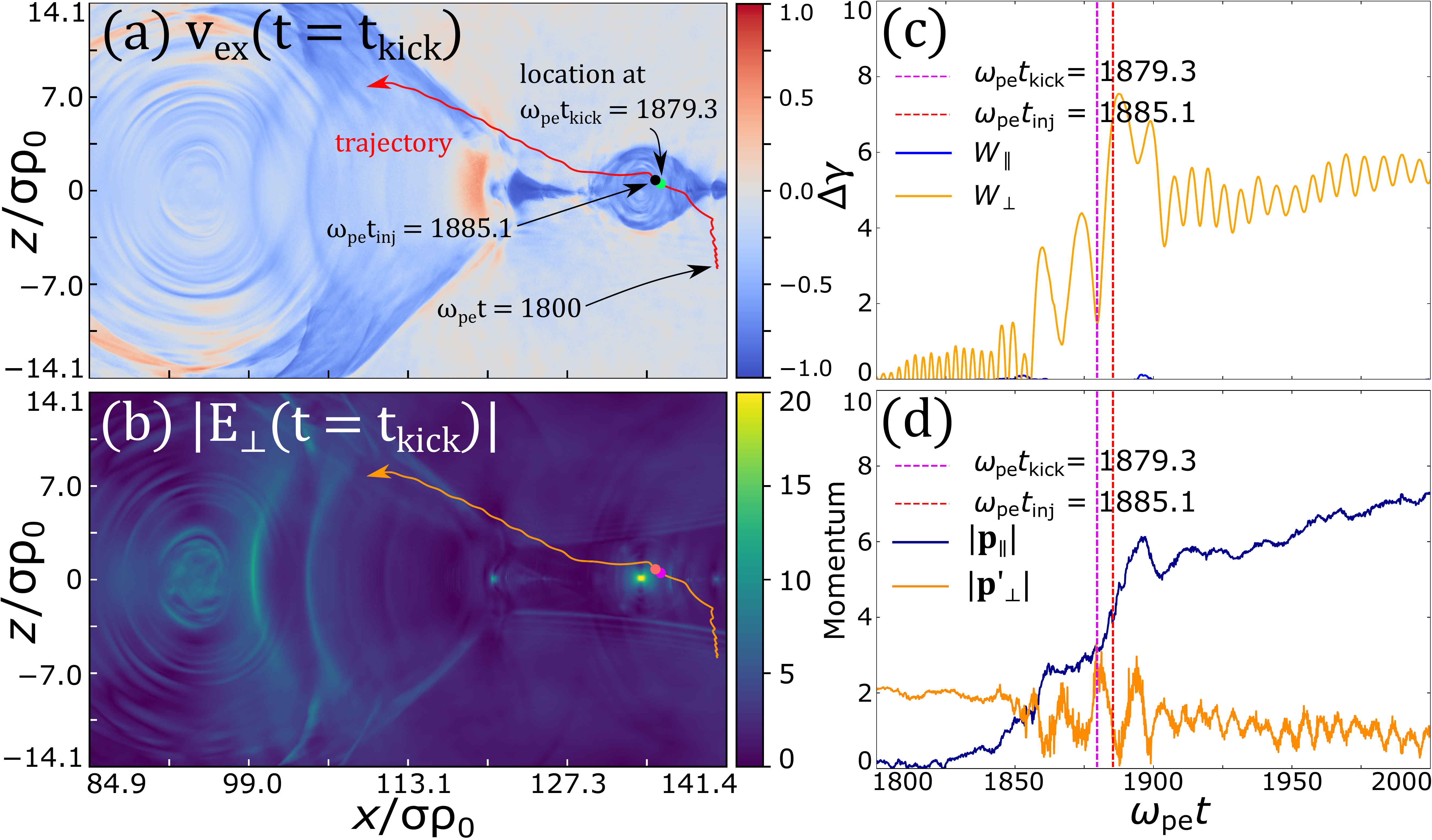}
    \caption{Example of Fermi acceleration, formatted similar to Figure~\ref{fig:direct_accel_example}.}
    \label{fig:Fermi_accel_example}
\end{figure*}

\begin{figure*}[htp!]
    \centering
    \includegraphics[width = 17.7cm]{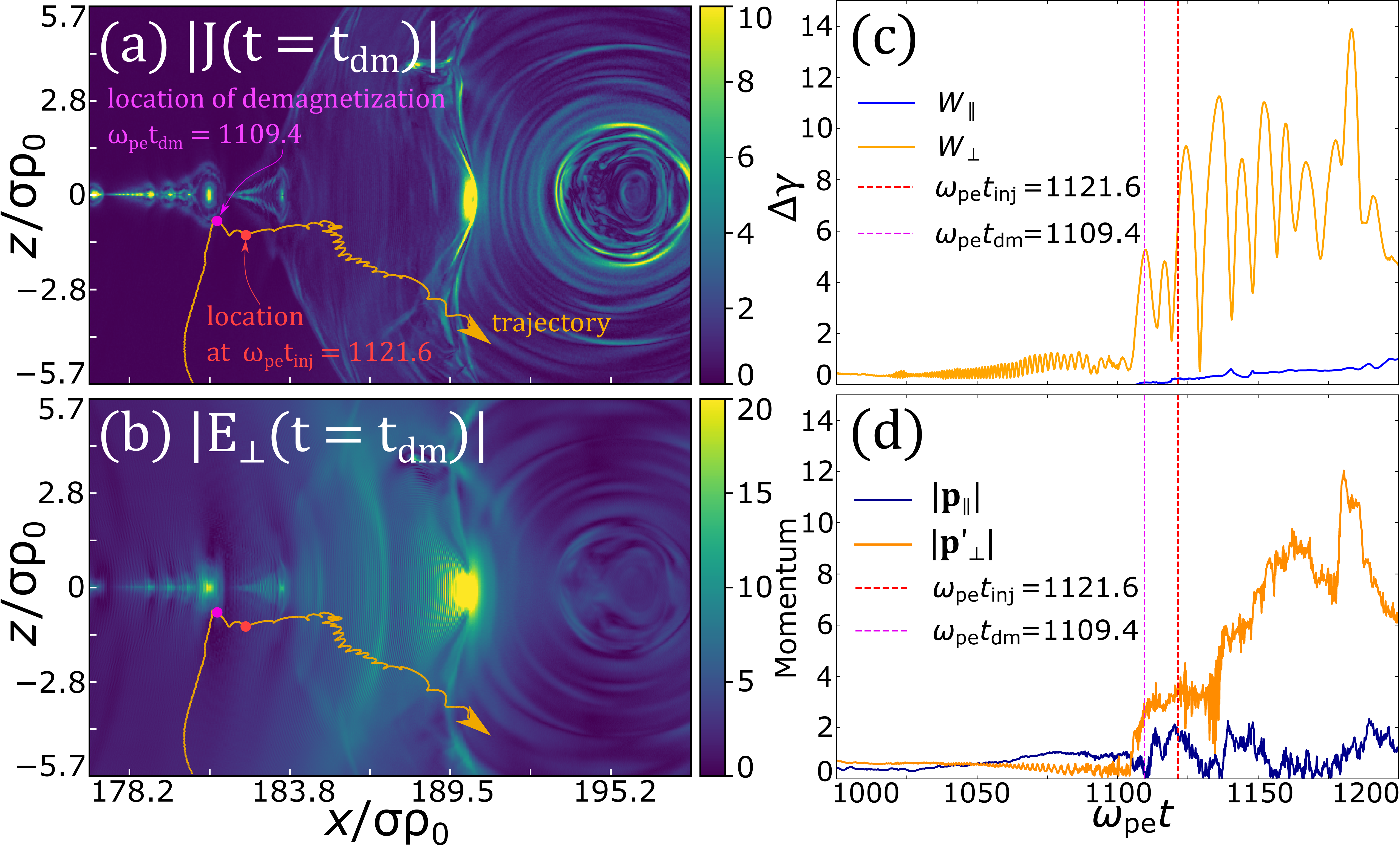}
    \caption{Example of pickup acceleration, formatted similar to Figure~\ref{fig:direct_accel_example}.}
    \label{fig:pickup_accel_example}
\end{figure*}

Before presenting the main statistical results, we show a representative example for each of the three injection mechanisms (see Figures~\ref{fig:direct_accel_example}-\ref{fig:pickup_accel_example}). Each of these examples presents the corresponding physical context within which such particle motion takes place (panels~a, b) and how the relevant quantities ($W_\parallel, W_\perp, \lvert \textbf{p}_\parallel \rvert,  \lvert \textbf{p}_\perp' \rvert$) vary with time (panels~c, d). 

Figure~\ref{fig:direct_accel_example} shows an example of particle injection by the reconnection electric field for a strong guide field of~$b_g = 1.0$ ($\ell_x = 289.6$).\footnote{We show the case of~$b_g = 1.0$ because injection mechanisms involving~$\textbf{E}_\perp$ are suppressed, and the~$E_\parallel$ region is significantly larger than in the weak guide field case \citep{Liu2020}, making a clearer picture of~$\textbf{E}_\parallel$ acceleration with fewer complications.} Once the electron enters the diffusion region (around~$\omega_{\rm pe}t=820$), it is subject to the reconnection electric field~$E_{\rm rec} \simeq E_y \simeq E_\parallel$, which continuously and steadily accelerates the particle from nonrelativistic ($\gamma \sim 1$) to injection ($\gamma_{\rm inj} \simeq 13$) and eventually to ultrarelativistic (e.g., $\gamma \sim 175$) energies. Once it passes through the vicinity of the X-point, the electron becomes caught by the reconnected (downstream) magnetic field on the right of the X-point (c.f., Figure~\ref{fig:accel_cartoons}b).

\begin{figure*}[htp!]
    \centering
    \includegraphics[width = 16.9cm]{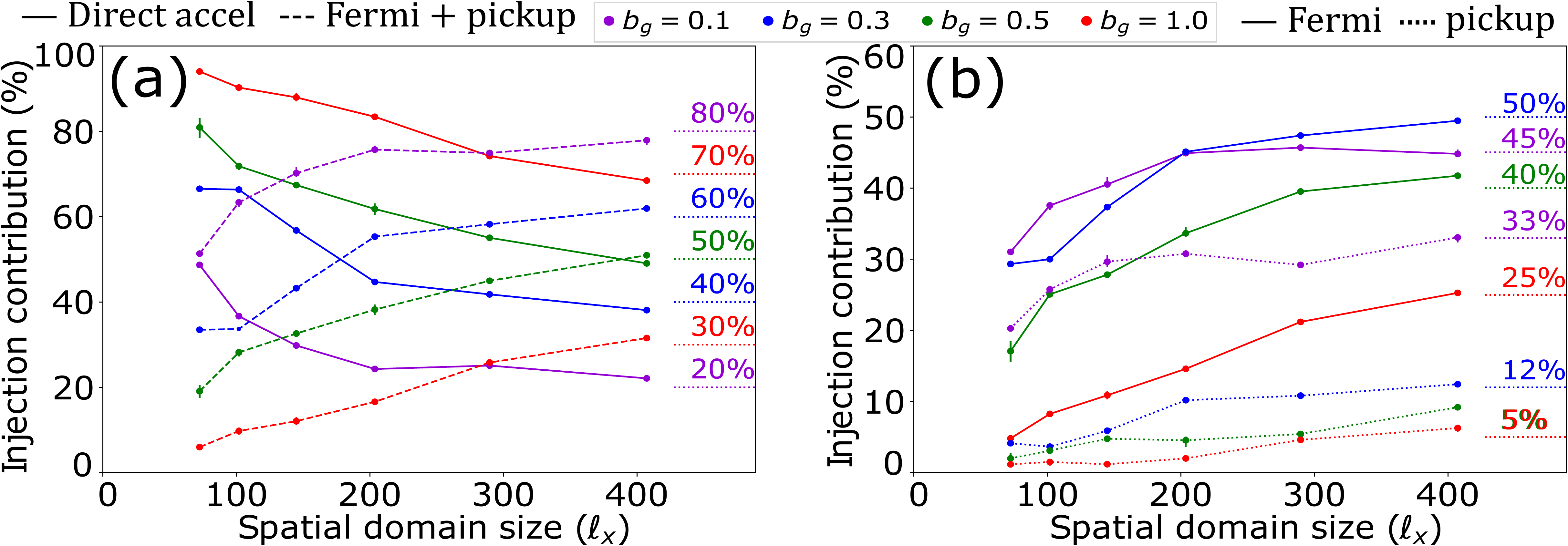}
    \caption{Cumulative percentage of injected electrons that are injected by each particle acceleration mechanism by~$t = \tau_f$, as defined by Eq.~(\ref{association2}). (a): Decomposition between particles injected by~$W_\parallel$ (solid) and~$W_\perp$ (dashed). (b): Decomposition between Fermi-injected particles (solid) and pickup-injected particles (dotted), as determined by Eq.~(\ref{association2}). Limiting values are labeled for visual aid.}
    \label{fig:injection_contributions}
\end{figure*}

Figure~\ref{fig:Fermi_accel_example} shows an example\footnote{We show the case of~$b_g = 0.1$ because Fermi kicks are considerably rarer for stronger guide fields (see Figure~\ref{fig:injection_contributions} for details).} of an electron accelerated through a Fermi kick
before it reaches~$\gamma_{\rm inj}$ inside a magnetic island for~$b_g = 0.1$ and~$\ell_x = 289.6$. Panel~(a) shows a color map of~$\textbf{v}_{ex}$ (i.e., the $x$-component of the electron flow velocity) and panel~(b) shows a color map of~$\lvert \textbf{E}_\perp \rvert$ at the time~$\omega_{\rm pe} t_{\rm kick} = 1879.3$, when the particle receives a Fermi kick. The particle begins in the upstream and approaches a reconnection outflow, where it is kicked by curved field lines. When the kick begins, the particle crosses the~$z = 0$ midplane and is suddenly energized by~$\Delta \gamma \sim 6$ via~$W_\perp$, causing it to surpass~$\gamma_{\rm inj}$. Simultaneously, the particle gains substantial parallel momentum~$\lvert \textbf{p}_\parallel \rvert$ and maintains its small gyroradius (c.f., Figure~\ref{fig:accel_cartoons}c). While not shown, the~$\textbf{E} \times \textbf{B}$ drift velocity rises from~$\sim 0.05 \,c$ to~$\sim 0.80 \,c$ over the interval~$\omega_{\rm pe}t \in [1800, 1879.3]$.

Lastly, an example\footnote{We show~$b_g = 0.1$ because pickup acceleration is weakened for stronger guide fields (see Figure~\ref{fig:injection_contributions} for details).} of the pickup process accelerating an electron during injection is shown in Figure~\ref{fig:pickup_accel_example} for $b_g = 0.1$ and~$\ell_x = 289.6$. At~$\omega_{\rm pe}t_{\rm dm} = 1109.4$, the particle experiences a sharp magnetic field change (with also low magnetic field strength) at the edge of an incoming plasmoid and subsequently becomes demagnetized (hence the subscript~``$\rm dm$"). Soon thereafter, the particle finds itself immersed in a relativistic, collisionless, magnetized plasma moving rapidly in the +$x$ direction. In contrast to Fermi acceleration, the particle does not cross the~$z = 0$ midplane. The particle is going through non-adiabatic motion as its magnetic moment~$\mu' \equiv \lvert \textbf{p}_\perp'\rvert ^2/2\lvert \textbf{B}' \rvert$ loses its invariance upon crossing the separatrix (not shown). Accordingly, the particle's gyroradius increases (c.f., Figure~\ref{fig:accel_cartoons}d). Simultaneously, the perpendicular electric field accelerates the particle, which begins to oscillate rapidly before joining a larger magnetic island. 

From these examples, we can clearly see that the particle trajectories match up reasonably well with each model drawn in Figure~\ref{fig:accel_cartoons}. Furthermore, the energy and momentum gains align with our expectations laid out in Eq.~(\ref{association2}). In the following subsection, we will apply Eq.~(\ref{association2}) to an ensemble of particles, which will generate for us a statistical understanding of how each mechanism contributes to particle injection across the $b_g$-$\ell_x$ scan.

\begin{figure*}[htp!]
    \centering
    \includegraphics[width = 16.8cm]{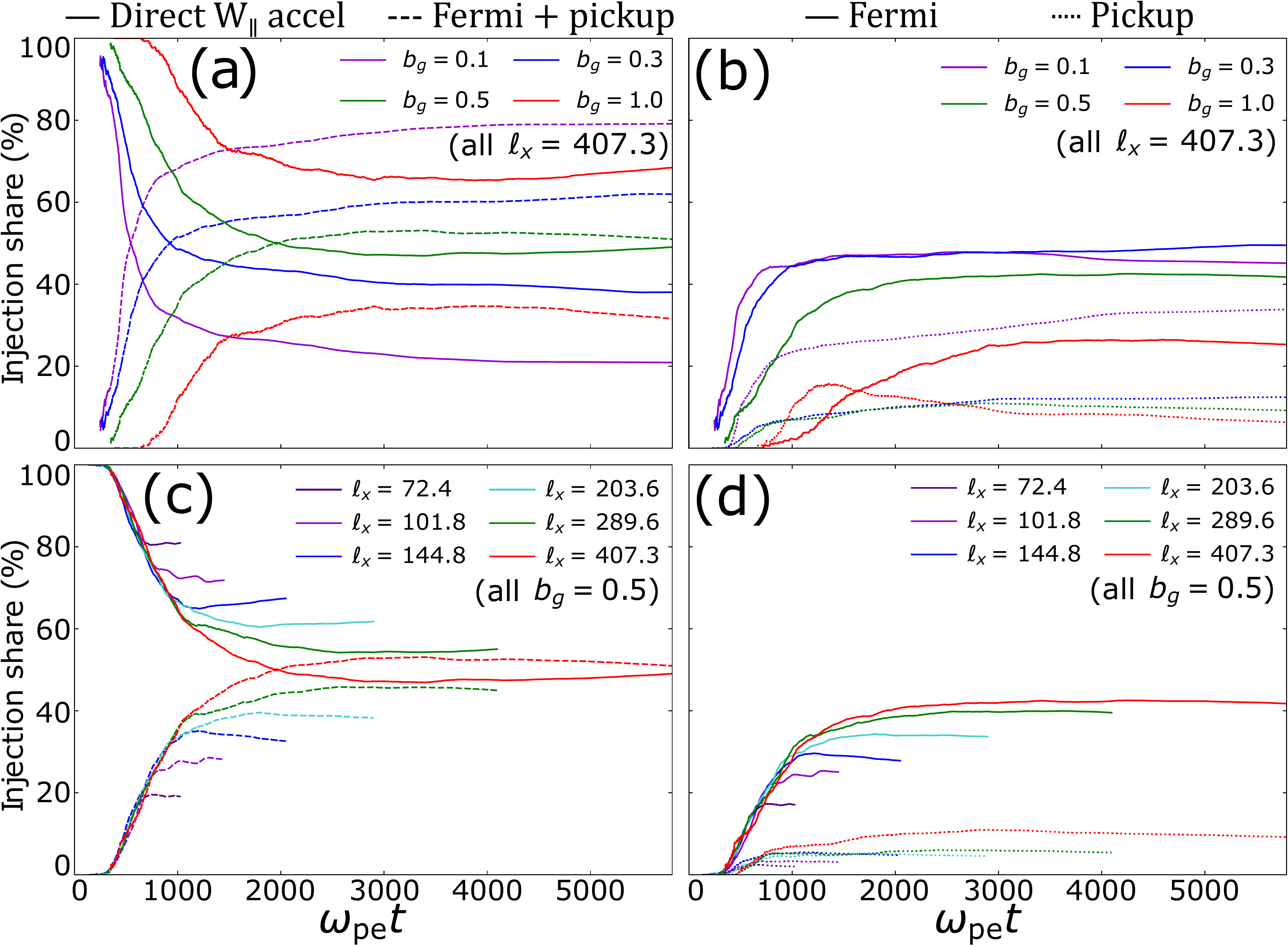}
    \caption{Evolution of cumulative particle injection shares. Panels~(a, c) show shares of particles injected primarily by direct acceleration (solid lines) versus the combined shares of Fermi and pickup acceleration (dashed lines). Panels~(b, d) contrast Fermi (solid) and pickup (dotted) injection shares. Panels~(a, b) vary guide-field strength at a constant domain size~$\ell_x = 407.3$ and~(c, d) vary domain size at  a constant guide-field strength~$b_g = 0.5$. A few lines are cut at early times due to small sample sizes, i.e., where very few particles are injected.}
    \label{fig:Ninj_vs_time}
\end{figure*}

\vspace{1cm}

\subsubsection{Statistical results}

Figure~\ref{fig:Ninj_vs_time} shows the statistical results on particle injection mechanisms during simulations for a series of guide-field strengths and domain sizes. The vertical axes are the shares of each injection mechanism. Panels~(a, c) suggest that~$W_\parallel$ acceleration dominates the injection process at early times, with its share decaying as time proceeds. For weak ($b_g = 0.1$, $0.3$) and moderate ($b_g = 0.5$) guide fields, the combined injection shares of the two $E_\perp$ (i.e., Fermi and pickup) injection mechanisms overtake that of direct ($E_\parallel$) acceleration at $\omega_{\rm pe}t \simeq 530$, $1020$, and~$1570$ respectively, whereas for a strong guide field ($b_g = 1.0$), direct acceleration remains dominant at all times. During the early injection phase, when~$W_\parallel$ is dominant (i.e., its share is $>50\%$), $W_\parallel$ injects only~$3\%$, $7\%$, and $20\%$ of the total, final injected population for~$b_g = 0.1$, $0.3$, and~$0.5$, respectively. Panels~(c, d) show that larger domains improve the Fermi and pickup injection shares, allowing them to eventually overtake~$W_\parallel$ for weak/moderate guide fields. However, given that domain-size convergence is not yet clear for~$b_g = 1.0$, it remains an open possibility that the~$W_\parallel$ injection share will eventually be overtaken by the combined Fermi and pickup injection share even for this guide field, provided that a sufficiently large domain is used. We also find that injection shares saturate with time, with stronger guide fields and larger domain sizes delaying saturation. Therefore, similar to~$\eta_N$ and~$\eta_E$ [Figure~\ref{fig:efficiencies}, panels~(a, b)], the injection saturation timescale appears to depend on the macroscopic dynamical timescale ($\sim L_x/c$), suggesting that the time of injection saturation is delayed for longer reconnection current sheets. This puts forward a picture that the time required to achieve the limiting injection mechanism partition depends on the length of the current sheet.

Particle injection results across the entire $\ell_x$-$b_g$ parameter scan at two light-crossing times are shown in Figure~\ref{fig:injection_contributions}. Panel~(a) shows the relative contributions to particle injection of direct acceleration ($W_\parallel$) versus combined Fermi and pickup acceleration ($W_\perp$), and panel~(b) decomposes the~$W_\perp$-injected particles into Fermi and pickup acceleration. Our findings can be summarized as follows. First, the share of electrons injected by pickup acceleration declines monotonically with increasing guide-field strength, with a notable sharp drop from~$\sim 33\%$ at~$b_g = 0.1$ to~$\sim 12\%$ at~$b_g = 0.3$, reaching~$\sim 5\%$ at~$b_g = 1.0$. Second, the Fermi injection share remains relatively constant at~$\sim 50\%$ for weak guide fields~$b_g = 0.1, 0.3$ and declines for stronger guide fields, down to~$\sim 25\%$ for the largest~$b_g = 1.0$ run. Third, the~$W_\parallel$ injection share increases substantially with increasing guide-field strength from~$20\%$ at~$b_g = 0.1$ to~$\sim 70\%$ at~$b_g = 1.0$.\footnote{However, in terms of total particle injection, $W_\parallel$ actually slightly declines as~$b_g$ strengthens. This can be seen as follows. Let us denote~$N_{\parallel}$ as the number of $W_\parallel$-injected particles and~$f_{\parallel}$ as the share of~$W_\parallel$-injected particles. Then, $N_{\parallel} = f_{\parallel} \eta_N N_{\rm ds}$. From~$b_g = 0.1$ to~$1.0$, we see that (a)~$f_{\parallel}(\tau_f)$ increases by a factor of~$\sim 3.5$ (Figure~\ref{fig:injection_contributions}a), (b)~$\eta_N(\tau_f)$ declines by a factor of~$\sim 2.5$ (Figure~\ref{fig:accel_efficiencies_tf}a), and (c)~$N_{\rm ds}(\tau_f)$ declines by a factor of~$\sim 1.8$ (not shown). Therefore, $W_\parallel$ actually injects fewer particles (by a factor of~$\sim 1.3$) when the guide field is strengthened from~$b_g = 0.1$ to~$1.0$.\label{fn:overall_wpara_inj}}

Increasing the domain size raises the injection shares of both~$W_\perp$ mechanisms, while the injection share due to~$W_\parallel$ decreases. For weaker guide fields, $b_g = 0.1, 0.3$, the injection shares appear to converge as~$\ell_x$ increases. However, for stronger guide fields~$b_g = 0.5, 1.0$, injection share convergence does not appear to be established, suggesting that stronger guide fields demand larger simulation domains to obtain the limiting, convergent contributions to particle injection as~$\ell_x$ increases (Figure~\ref{fig:injection_contributions}). In the following paragraphs, we discuss some possible explanations for these trends.

Let us begin by explaining how varying guide-field strength affects particle injection. Recall that in relativistic ($\sigma_h \gg 1$) reconnection, strengthening the guide field~$b_g$ reduces the in-plane Alfv\'en speed~$v_{Ax}$ \citep{Liu2015,Werner2017,Guo2020,Uzdensky2022}.\footnote{In particular, $v_{Ax} = c\,\{\sigma/[1 + \sigma(1 + b_g^2)]\}^{1/2}$, which simplifies to $v_{Ax} = (1 + b_g^2)^{-1/2}$ (normalized to~$c$) when $\sigma \gg 1$ \citep{Uzdensky2022}.} As we will see in the subsequent paragraphs, this reduction in~$v_{Ax}$ has implications for every injection mechanism.

\begin{figure*}[htp!]
    \centering
    \includegraphics[width = 18cm]{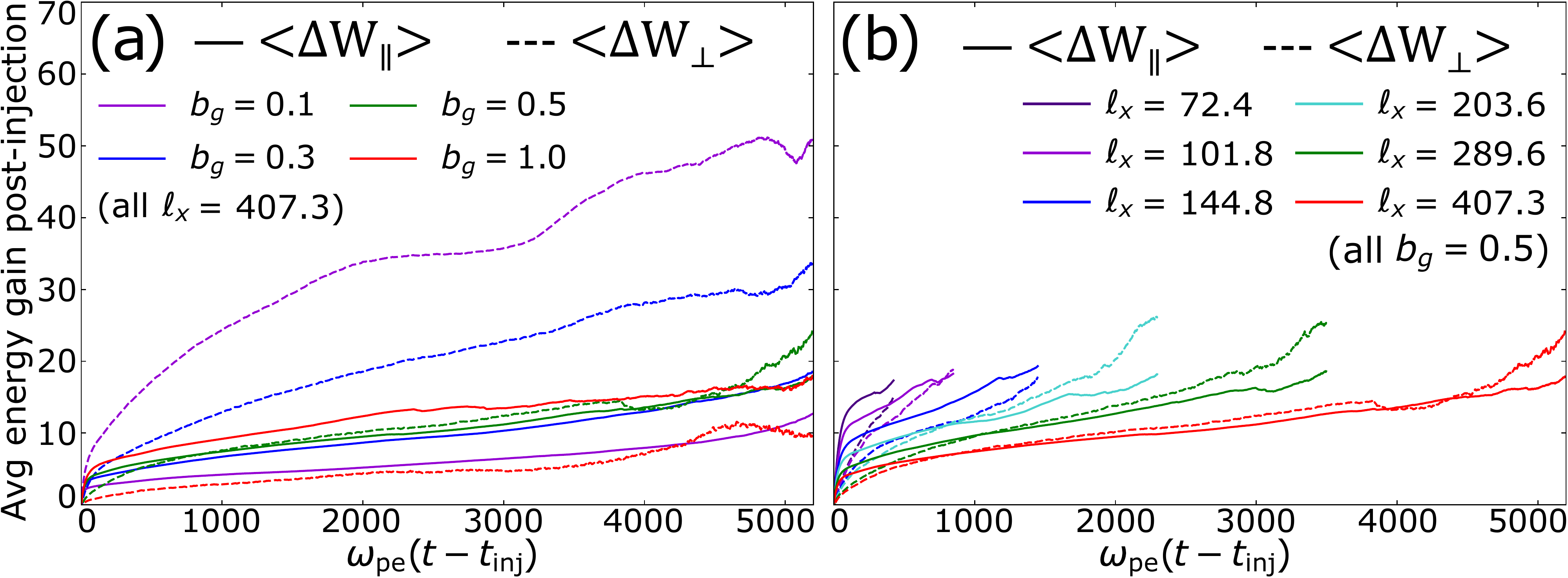}
    \caption{Evolution of the average electron kinetic energy gained from~$W_\parallel$ (solid lines) and~$W_\perp$ (dashed lines) after injection (i.e., the horizontal axis is offset by the injection time~$t_{\rm inj}$). Each line is cut short by~$\sim 600 \omega_{\rm pe}t$, as, beyond this limit, the averaging becomes unreliable due to a significant drop in sample size (i.e., the displayed domain is~$\omega_{\rm pe}(t - t_{\rm inj}) \in [0, 5160]$ instead of the entire~$[0, 5760]$). (a): Results for varying guide-field strength at a fixed domain size~$\ell_x = 407.3$. (b): Results for varying domain size at a fixed  guide-field strength~$b_g = 0.5$.}
    \label{W_vs_time_postinj}
\end{figure*}

\begin{figure}[htp!]
    \centering
    \includegraphics[width = 8.5cm]{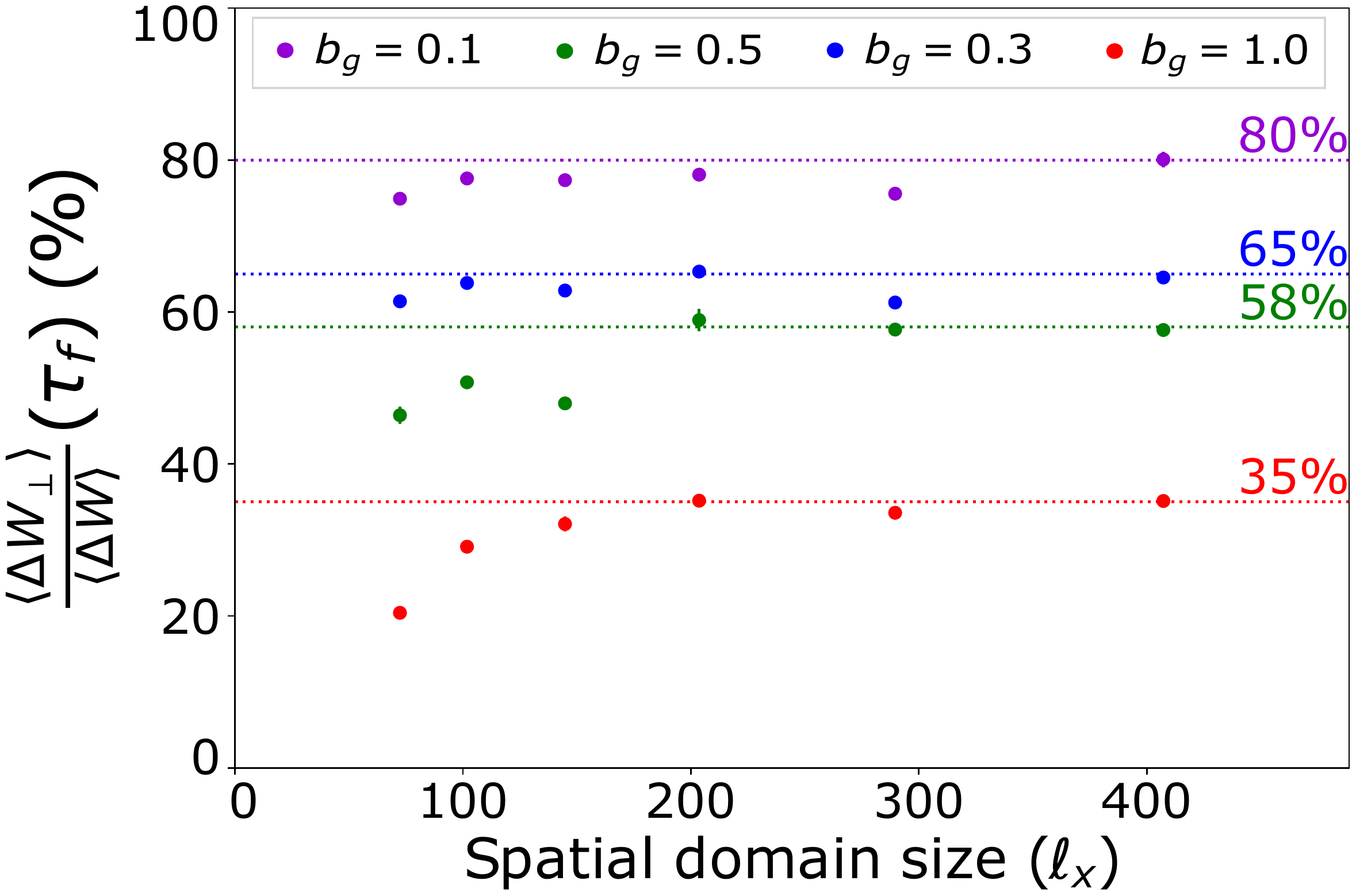}
    \caption{Average contribution of~$W_\perp$ to high-energy ($\gamma > \gamma_{\rm inj}$) particle acceleration at~$t = \tau_f$ for different domain sizes and guide fields. The corresponding contributions from~$W_\parallel$ are the complement of each data point. Horizontal lines are shown for visual aid.}
    \label{fig:postinj_results}
\end{figure}

Particle injection by the pickup process depends upon the in-plane Alfv\'en speed~$v_{Ax}$ (Figure~\ref{fig:accel_cartoons}d), so we may expect any energy gains from pickup to be reduced as~$b_g$ strengthens. Let us explore whether this can explain the sharp drop in the pickup injection share observed between~$b_g = 0.1$ and $b_g = 0.3$ (Figure~\ref{fig:injection_contributions}b). We assume that the energization from pickup acceleration is given by
$$ W_{\rm pickup} \simeq \gamma_{Ax} - 1 = \frac{1}{\sqrt{1 - v_{Ax}^2}} - 1 = \sqrt{1 + \frac{\sigma}{1 + \sigma b_g^2}} - 1. $$
Plugging in~$\sigma = 50$ and~$b_g = 0.1$, $0.3$, $0.5$, $1.0$ yields~$W_{\rm pickup} \simeq 4.9$, $2.2$, $1.2$, $0.4$ respectively. Therefore, even a slight variation of~$b_g$ from~$0.1$ to~$0.3$ results in a sharp (over~$50\%$) decline in the work done by the pickup process. This may partially explain why pickup injection shares fall drastically over this variation. As a point of speculation, we note that raising~$b_g$ from~$0.1$ to~$0.3$ could also cause more particles to remain adiabatic around typical demagnetization regions (e.g., the mid-plane or plasmoids).

Let us now provide a few comments about the dependence of~$W_\parallel$ injection on~$b_g$. The reduction of the in-plane Alfv\'en speed weakens the reconnection electric field~$E_{\rm rec} \simeq E_\parallel$, leading to a reduction in~$W_\parallel$ injection \citep{Werner2017,Werner2018,Uzdensky2022}. Indeed, as mentioned in Footnote \ref{fn:overall_wpara_inj}, we observe a modest suppression of the~$W_\parallel$ injection with increasing~$b_g$. However, we note that the suppression of the~$W_\parallel$ injection is much less pronounced than the suppression of the~$W_\perp$ injection, where a decreasing injection share is accompanied by a reduction in~$\eta_N$ as well.

Finally, let us discuss the trends shown in Figure~\ref{fig:injection_contributions} with increasing domain size~$\ell_x$. First, because the initial current sheet spans the entire domain ($x \in [0, \ell_x]$), the rate at which it becomes occupied with plasmoids is independent of domain size. Therefore, simulations with larger~$\ell_x$ (and hence a longer running time~$\tau_f$) will have a greater fraction of their total running time in the stage where particles are accelerated somewhere other than the original, primary current sheet---the main location of particle injection by~$E_{\rm rec} \simeq E_\parallel$. Meanwhile, the opportunities for Fermi and pickup injection are less suppressed with time and therefore are less suppressed with increasing~$\ell_x$.

In summary, we have applied Eq.~(\ref{association2}) to a sample of~$\sim 400,000$ tracer particles, yielding a statistical picture of how each mechanism contributes to particle injection. We have plotted the contributions of each mechanism for various guide-field strengths and domain sizes and have attempted to describe some of the observed trends.

\subsection{Post-injection particle acceleration} \label{ss:postinj_results}

The particle acceleration mechanism(s) responsible for the high-energy nonthermal particle distribution ($\gamma \gtrsim \gamma_{\rm inj}$) can be diagnosed by evaluating the average work done by parallel ($\langle \Delta W_\parallel \rangle (t)$) and perpendicular ($\langle \Delta W_\perp \rangle (t)$) electric fields, where~$\langle \ \rangle$ indicates an average over all injected tracer particles and~$\Delta$ indicates the additional energization beyond $\gamma_{\rm inj}$, i.e., after~$t = t_{\rm inj}$ for a given particle.

First, we show the evolution of~$\langle \Delta W_\parallel \rangle$ and~$\langle \Delta W_\perp \rangle$ for different guide-field strengths at a fixed system size~$\ell_x = 407.3$ (Figure~\ref{W_vs_time_postinj}a) and for different system sizes at a fixed guide field~$b_g = 0.5$ (Figure~\ref{W_vs_time_postinj}b).

When increasing system size with a fixed~$b_g$, we see that the average final-time energy gains by~$\langle \Delta W_\parallel \rangle$ for each system size are similar, and the energization rate is inversely proportional to $\ell_x$. This suggests that the timescale of post-injection energy gain by~$E_\parallel$ is controlled by the length of the current sheet~$\propto \ell_x$ (Figure~\ref{W_vs_time_postinj}b). In contrast, final-time energy gains by $\langle \Delta W_\perp \rangle$ appear to improve for larger domains. However, as we will show below, this improvement may quickly saturate with increasing domain size.

Figure~\ref{fig:postinj_results} shows the average percentage contribution from~$W_\perp$ to total high-energy ($\gamma \gtrsim \gamma_{\rm inj}$) particle acceleration at final times~$t = \tau_f$ for every simulation across the parameter scan. Variation with spatial domain size appears to saturate rapidly, while the trend with guide-field strength is robust and dramatic, with~$W_\perp$ dropping from being dominant ($\sim 80\%$) at a weak~$b_g = 0.1$ guide field to only~$\sim 35\%$ at a strong~$b_g = 1.0$ guide field. Note that these fractions may contribute more favorably to~$\langle \Delta W_\perp \rangle$ if longer simulations are ran (i.e., we use~$\tau_f > 2L_x/c$).

\section{Discussion} \label{sec:5_discussion}

\subsection{Astrophysical applications} \label{ss:applications}
A detailed understanding of nonthermal particle acceleration is essential for characterizing nonthermal radiation signatures in astrophysical sources. Our study complements recent work in radiative relativistic magnetic reconnection that has been a promising start for connecting particle-in-cell simulations to observational high-energy astrophysics, both in terms of theory \citep{Jaroschek2009, Uzdensky2011, Uzdensky2016, Nalewajko2018, Macias2019, Hakobyan2019, Werner2019, Sironi2020, Mehlhaff2020, Mehlhaff2021, Nattila2021} and application \citep{Uzdensky_etal-2011, Cerutti2012, Cerutti2013, Cerutti2014, Cerutti2014b, Beloborodov2017, Philippov2019, Schoeffler2019, Zhang2022, Hakobyan2022, Sridhar2022, Chen2022}.

This paper takes an important step in advancing the research program of using potentially-observed (via photon spectra and light curves) quantities (e.g., the power-law index~$p$, the low-energy cutoff~$\gamma_{\rm inj}$, the high-energy cutoff~$\gamma_c$, flare duration, flare intensity) to reverse-engineer the unobservable, characteristic parameters of the system (e.g., $\sigma$, $b_g$, $\ell_x$). In particular, we have made progress by obtaining relationships between the observable quantities ($p$, $\gamma_{\rm inj}$, $\gamma_c$, $\eta_N$, $\eta_E$) and the unobservable quantities ($b_g$, $\ell_x$) from first-principles.

Moreover, establishing convergence of these parameters with increasing domain size (spatial and temporal) is particularly important for astrophysical applications, as the convergence of a given parameter indicates that its value may remain stable when extrapolated to the very large scales of real astrophysical systems. Indeed, we have found convergence of~$p$, $\gamma_{\rm inj}$, $\eta_N$, and~$\eta_E$ with increasing domain size (i.e., Figures~\ref{spectra_results}a, b; \ref{fig:gamma_inj_fitted}, \ref{fig:accel_efficiencies_tf}a, b), suggesting that these results can be extrapolated to astrophysical scales of space and time.

Let us now discuss some potential applications for some of these quantities. First, the evolution of the energy efficiency~$\eta_E(t)$ is valuable for analyses and inferences of the energy content in astrophysical systems. The initial transient rise in~$\eta_E$ demonstrates that relativistic reconnection can rapidly (i.e., on a timescale independent of~$\ell_x$) convert large amounts of magnetic energy into nonthermal particle acceleration in highly magnetized plasmas. On the other hand, having the saturation timescale of~$\eta_E$ be directly proportional to the length~$\ell_x$ of the initial current sheet suggests that larger systems have longer timescales of active plasma dynamics (e.g., plasmoid mergers), which may be relevant to observations. Lastly, $\eta_E(b_g)$ helps us understand how the guide-field strength suppresses the efficiency of energy conversion due to reconnection. 

Second, the power-law index of photon spectra can often be used to infer the power-law index of particle spectra if several simplifying assumptions can be justified, allowing direct comparison to observations (e.g., see \citealt{Werner2021}). Knowing~$p$ (along with~$\gamma_{\rm inj}$ and~$\gamma_c$) also provides a constraint on the total energy content available in the nonthermal particle spectrum.

Since the high-energy cutoff~$\gamma_c$ (Figure~\ref{spectra_results}c) appears not to converge with increasing system size~$\ell_x$ for weak-to-moderate guide fields, this potentially observable quantity cannot eliminate the degeneracy of, e.g., the power-law index $p = p(b_g, \sigma)$. However, seeing that~$\gamma_{\rm inj}$ does appear to converge with increasing~$\ell_x$ for these cases, it may have a well-defined dependence $\gamma_{\rm inj} = \gamma_{\rm inj}(b_g, \sigma)$ for asymptotically large~$\ell_x$. Therefore, if~$p$ and~$\gamma_{\rm inj}$ can be deduced from observation of some astrophysical flaring event, then both~$b_g$ and~$\sigma$ can in principle be inferred. This opens up a possibility for~$\gamma_{\rm inj}$, a relatively new and unexplored quantity, to be a crucial diagnostic, along with~$p$, for inferring the underlying system parameters from the observed spectra.

\vspace{1cm}
\subsection{Comparisons with previous work} \label{ss:comparisons}
In \citealt{Ball2019}, a transrelativistic ($\sigma = 0.3$) proton-electron plasma with guide fields~$b_g = 0.1$ and $0.3$ is used. The authors find that primary and secondary (in the case of~$b_g = 0.1$) X-points are dominant sites of electron injection owing to~$W_\parallel$, and that electron spectra above~$\sigma/2$ are dominated by electrons injected near X-points. Our work concludes differently in a few crucial aspects. We find that overall particle injection for these weak guide fields is completely dominated by~$W_\perp$ mechanisms (Fermi and pickup acceleration), despite~$W_\parallel$ injecting particles first, presumably in primary X-points. Plasmoids are present in all of our simulations, yet we do not see a substantial improvement in~$W_\parallel$ injection shares at later times when secondary X-points arise (Figures~\ref{fig:Ninj_vs_time} and~\ref{fig:injection_contributions}). 
A follow-up study to \citealt{Ball2019} was done by \citealt{Kilian2020}, which has a similar simulation setup (proton-electron transrelativistic plasma) and focuses on the case of~$b_g = 0.1$. The authors find that while~$W_\parallel$ injects the first few particles, it is~$W_\perp$ injection that dominates in the longer term. They also find that~$W_\perp$ dominates post-injection acceleration. These conclusions are in close agreement with our study (Figures~\ref{fig:Ninj_vs_time} and~\ref{fig:injection_contributions}), suggesting this conclusion applies to several different plasma regimes.

In \citealt{Guo2019}, it is claimed that the formation of power-law distributions does not rely on non-ideal MHD electric fields. Instead, ``motional" (i.e., induced by bulk plasma motion) ideal electric fields~$\textbf{E}_m \equiv -\textbf{u} \times \textbf{B}$ are responsible for power-law formation. While we have not attempted to distinguish between ideal and non-ideal MHD electric fields in this study, we have evaluated the contributions of~$W_\perp$ and~$W_\parallel$ to secondary~$\gamma > \gamma_{\rm inj}$ particle acceleration, which can serve as proxies for ideal and non-ideal electric fields respectively. It has been argued in~\citealt{Uzdensky2022}, however, that highly energetic particles ($\gamma \gg \gamma_{\rm inj}$) will, in general, have characteristic gyroradii~$\rho(\gamma) \equiv \gamma \rho_0 \gg \delta$, where~$\delta$ is the thickness of the smallest elementary current layers, making the contributions of non-ideal electric field components to post-injection acceleration not well-defined (and hence, evaluating~$W_\parallel$ and~$W_\perp$ may not be very relevant for understanding particle acceleration). Nevertheless, our results indicate that perpendicular electric fields play an important role in accelerating particles after they are injected, although their relative importance declines with increasing guide-field strength (Figure~\ref{fig:postinj_results}).

In \citealt{Sironi2022} (henceforth S22), the non-ideal electric field~$\textbf{E}_n$ is proposed to solve the injection problem in relativistic reconnection, i.e., the question of how particles from a thermal upstream are accelerated to the high energies entering the power-law tail. The author approximates~$\textbf{E}_n$ by the local electric field in regions where~$E > B$ when~$b_g \lesssim \eta_{\rm rec}$ and by~$E_\parallel \sim \eta_{\rm rec}B_0$ when~$b_g \gtrsim \eta_{\rm rec}$. S22 finds this conclusion for both 2D and 3D simulations. From this conclusion, we would expect mechanisms that use the motional electric field~$\textbf{E}_m = \textbf{E} - \textbf{E}_n$---such as Fermi and pickup---to contribute negligibly to particle injection. However, we find that Fermi and pickup play a dominant role in injecting particles for~$b_g = 0.1$, $0.3$, and even a non-negligible role when~$b_g = 1.0$ (Section~\ref{ss:inj_results}). There is also a Comment by \citealt{Guo2022_comment} that challenges the conclusions of S22 directly, particularly for the case without a guide field, and employs a similar methodology to our work. The Comment finds that the amount of time particles spend in~$E > B$ regions is, in general, insufficient for injection. In particular, most of the energy gained by~$E > B$ particles is done outside~$E > B$ regions. Furthermore, the Comment finds that the average energy gained by~$E > B$ particles \textit{before} encountering~$E > B$ is comparable to the average energy gained within~$E > B$ regions, suggesting that~$E > B$ regions are not unique in pre-accelerating particles.

In \citealt{Hoshino2022} (henceforth H22), a systematic investigation of the efficiency of nonthermal particle acceleration in magnetic reconnection is conducted, where simulations are 2D, using a pair plasma, are without guide field, and run over a scan of eight values of~$\sigma$ (from~$0.02$ to~$\simeq 63$) while maintaining~$\sigma_h \equiv \sigma/\theta = 2$ fixed. Acceleration efficiencies~$\varepsilon_{\rm den}$, $\varepsilon_{\rm ene}$ in H22 are defined similarly to~$\eta_N$, $\eta_E$ in our work (Section~\ref{ss:accel_efficiency}). Our largest run with~$b_g = 0.1$ may be close enough to~$b_g = 0$ for a direct comparison of efficiencies. In particular, when comparing the run with ($\sigma \simeq 63$, $\theta \simeq 32$) in H22 to our run (i.e., with $\sigma = 50$, $\theta = 0.25$), we find strong agreement between energy efficiencies ($\eta_E \simeq \varepsilon_{\rm ene} \simeq 90\%$), but a significant difference between number efficiencies ($\eta_N \simeq 40\%, \ \varepsilon_{\rm den} \simeq 60\%$) (Figure~\ref{fig:accel_efficiencies_tf}). The discrepancy between the number efficiencies in our studies tentatively suggests that, in the~$\sigma \gg 1$ regime of reconnection, a greater upstream temperature improves the efficiency of particle injection. However, a more thorough exploration of~$\eta_N(\theta)$ and~$\eta_E(\theta)$ is needed before any definitive conclusion can be drawn. To investigate the effect of the cold magnetization~$\sigma$, we can compare the ($\sigma \simeq 0.63$, $\theta \simeq 0.32$) run in H22 to our run. We find agreement between both efficiencies ($\eta_N \simeq \varepsilon_{\rm den} \simeq 40\%, \eta_E \simeq \varepsilon_{\rm ene} \simeq 90\%$), tentatively suggesting that they are independent of~$\sigma$ when~$\sigma \gtrsim 1$. However, similar to~$\theta$-dependence, a more thorough investigation of~$\eta_N(\sigma)$ and~$\eta_E(\sigma)$ is needed.

\vspace{1cm}
\subsection{Future computational work} \label{ss:future_work}
Let us outline prospects for future follow-up computational studies of reconnection-driven particle acceleration in a relativistic collisionless plasma. In particular, let us focus on particle injection. The criteria for characterizing injection mechanisms can be enhanced to incorporate particle location (relative to the relevant plasma and magnetic structures) into each condition. A similar approach has been done in the context of nonthermal particle acceleration \citep{Nalewajko2015}. To do this, we could develop automatic procedures that identify the relevant structures (electron diffusion regions, reconnection outflows, plasmoids, etc.) without bias. Such procedures could be similar to those already developed for identifying X-points \citep{Haggerty2017} and downstream regions \citep{Daughton2014}.

In order to have a more realistic and astrophysically-relevant picture of nonthermal particle acceleration, it will be necessary to increase the complexity of the problem in two major respects: increasing the dimensionality to three and incorporating the effects of radiation. 

In 3D simulations of the non-relativistic regime, particles may leak out from magnetic islands along chaotic field lines and explore multiple exhausts to improve the efficiency of Fermi acceleration\citep{Dahlin2017,Zhang2021a,Johnson2022}. In particular, in the weak guide field regime where Fermi acceleration is the strongest, the flux-rope kink instability disintegrates the magnetic flux ropes to facilitate particle transport \citep{Zhang2021a}. In the relativistic regime, however, this energetic-particle transport along field lines is limited by the speed of light~$c$, close to the reconnection outflow speed~$v_{Ax}$. Therefore, it is difficult for the energetic particles to explore multiple exhausts to increase the acceleration efficiency in 3D. This may explain why electron spectra have been found to be similar between 2D and 3D simulations in the relativistic regime \citep{Werner2017,Werner2021,Guo2021}.

Let us now turn to radiation, which is the second major source of complexity that must be incorporated into future studies. Inside small-scale diffusive regions, where rapid~$W_\parallel$ acceleration by~$E_{\rm rec} \sim E_\parallel$ takes place, the magnetic field is relatively weak, and therefore energy losses via synchrotron cooling are less significant, especially compared the timescales of acceleration \citep{Uzdensky_etal-2011,Cerutti2012}. Likewise, pickup processes may accelerate particles on similarly short timescales \citep{Sironi2020}. However, in large magnetic islands where a continual Fermi acceleration process may occur, the magnetic field strength is much stronger and acceleration occurs over a much longer duration, so losses via synchrotron cooling become significant \citep{Schoeffler2019,Werner2019,Zhdankin2020}. Therefore, investigations that account for the complexity of 3D effects and radiation will be essential ``next steps" toward a more complete and astrophysically-relevant description of particle acceleration.

\section{Conclusions} \label{sec:6_conclusions}

In this work, we performed an array of 2D particle-in-cell simulations of relativistic ($\sigma \equiv B_0^2/4\pi nmc^2 = 50$) collisionless magnetic reconnection, using the code \code{VPIC}. From these simulations, we had two primary goals. The first was to measure the quantities that characterize particle spectra, namely~$\gamma_{\rm inj}$, $\gamma_c$, and~$p$ (Section~\ref{ss:spectra}) and dependent quantities~$\eta_N, \ \eta_E$ that provide concrete notions of acceleration efficiency (Section~\ref{ss:accel_efficiency}), as functions of guide magnetic field strength and system size. The second goal was to evaluate the contributions of three nonthermal particle acceleration mechanisms to particle injection (Section~\ref{ss:inj_results}). Additionally, we also investigated the relative roles of parallel and perpendicular electric fields to high-energy particle acceleration (Section~\ref{ss:postinj_results}). To achieve these goals, we developed a new diagnostic to calculate the injection energy~$\gamma_{\rm inj}$. 

Our simulations span six domain sizes~$\ell_x$ and four normalized guide-field strengths~$b_g$, which we used to investigate convergence with increasing domain size and the guide-field dependence of the aforementioned quantities at the largest domains. We note that we have not performed analytical fits as a function of guide-field strength, as (in our view) there are not enough values in the~$b_g$ scan for such a fit to be reliable.

\subsection{Particle spectra}

\begin{enumerate}
    \item \textbf{Power-law index~$\boldsymbol{p}$:} Nonthermal power-law indices~$p$ increase with increased domain sizes at moderate~$\ell_x$, but then appear to converge at sizes above $\ell_x \gtrsim 200$ (Figure~\ref{spectra_results}a). This suggests that our simulation results can be extrapolated to extremely large, e.g., astrophysical, systems.
    
    We also find that stronger guide fields steepen the nonthermal spectra that emerge from relativistic magnetic reconnection. In particular, the asymptotic values of~$p$ in the $\ell_x \to \infty$ limit increase with~$b_g$; specifically, we find $p = 1.90$, $1.90$, $2.20$, and $2.80$ for~$b_g = 0.1, 0.3, 0.5, 1.0$, respectively.
    
    \item \textbf{Injection energy~$\boldsymbol{\gamma_{\rm inj}}$:} 
    The injection energy~$\gamma_{\rm inj}$ (the energy at which the power-law component begins) converges with increasing~$\ell_x$. We performed an analytical fitting to obtain the asymptotic injection energy~$\gamma_{\rm inj}'$. We find that this limiting injection energy, when normalized to the upstream magnetization~$\sigma$, has a moderate guide-field dependence, rising from $\gamma_{\rm inj}'/\sigma \sim 0.15$ for a weak guide field of~$b_g = 0.1$ to ~$\sim 0.30$ for a strong guide field of~$b_g = 1.0$.
    
    \item \textbf{High-energy cutoff~$\boldsymbol{\gamma_c}$:} 
    The high-energy cutoff $\gamma_c$ grows with increasing system size. Interestingly, the growth rate of~$\gamma_c$ with increased~$\ell_x$ weakens with strengthening~$b_g$, from approximately linear when~$b_g = 0.1$ to convergent when~$b_g = 1.0$, with sub-linear growth for intermediate guide fields. The ratio~$\gamma_c/\sigma$ ranges from~$\sim 3$ to~$\sim 15$ across the parameter scan (Figure~\ref{spectra_results}c).
    
    \item \textbf{Power-law extent~$\boldsymbol{\gamma_{c}/\gamma_{\rm inj}}$:} Given the convergent trend of~$\gamma_{\rm inj}$ with increasing~$\ell_x$, we expect the power-law extents (dynamic ranges) to increase with domain size in accordance with the growth trends of~$\gamma_c$ with increasing~$\ell_x$. They also shrink with an increased guide field. Over the entire parameter scan of~$b_g$ and~$\ell_x$, extents fall within 1-2 decades; for the largest domain, $\ell_x = 407.3$,  $\gamma_c/\gamma_{\rm inj}(b_g = 0.1) \sim 90$ and $\gamma_c/\gamma_{\rm inj}(b_g = 1.0) \sim 20$ (Figure~\ref{spectra_results}d).
\end{enumerate}

\subsection{Acceleration efficiency}

\begin{enumerate}
    \item \textbf{Time evolution of efficiencies:} We defined two notions of acceleration efficiency: the number efficiency~$\eta_N \equiv N_{\rm inj}/N_{\rm ds}$ [Eq. (\ref{equation:n_inj})] and the energy efficiency~$\eta_E \equiv E_{\rm inj}/E_{\rm ds}$ [Eq.~(\ref{equation:e_inj})]---i.e., the nonthermal (beyond $\gamma_{\rm inj}$) particles' share of the particle number and energy relative to the total particle number and energy of the downstream population.
    
    We find that, for weak and moderate guide fields~($b_g = 0.1, 0.3, 0.5$), both $\eta_N$ and~$\eta_E$ experience a rapid initial transient rise followed by a moderate fall, after which they climb again slowly to their final asymptotic values on longer timescales. As the length of the current sheet $\ell_x$ increases, the saturation timescale for each efficiency increases proportionally (Figure~\ref{fig:efficiencies}b, d). For the strong guide field~$b_g = 1.0$, however, the initial rises are much less pronounced and saturation takes a longer time (Figure~\ref{fig:efficiencies}a, c).

    \item \textbf{Efficiencies at the final time~$\boldsymbol{t = \tau_f}$:} 
    We find that~$\eta_N(\tau_f)$ and~$\eta_E(\tau_f)$ vary little with domain size, suggesting that saturation with increasing~$\ell_x$ has been reached. Specifically, $\eta_N(\tau_f)$ approaches~$\sim 35\%$ for weak guide fields~($b_g = 0.1, 0.3$), declining to~$\sim 15\%$ for a strong guide field~$b_g = 1.0$  (Figure~\ref{fig:accel_efficiencies_tf}a). The final energy efficiency~$\eta_E(\tau_f)$ approaches~$\sim 90\%$ for~$b_g = 0.1, 0.3$ and~$\sim 60\%$ for~$b_g = 1.0$ (Figure~\ref{fig:accel_efficiencies_tf}b).
\end{enumerate}

\subsection{Particle injection mechanisms}
\begin{enumerate} 
    \item \textbf{Time evolution of injection shares:} We find that direct~$W_\parallel$ acceleration dominates the injection process at early times, suggesting that the initial reconnection events inject particles first. However, the~$W_\parallel$ share decays as time (in terms of~$\omega_{\rm pe}^{-1}$) proceeds (Figure~\ref{fig:Ninj_vs_time}a, c) towards a saturation, such that for weak-to-moderate guide fields ($b_g = 0.1$, $0.3$, $0.5$) the combined injection shares of the two $W_\perp$ acceleration mechanisms (Fermi and pickup) eventually overtake that of direct acceleration. During the early~$W_\parallel$-dominant stage, $W_\parallel$ only injects~$3\%$, $7\%$, and $20\%$ of the total, final injected population for~$b_g = 0.1$, $0.3$, and $0.5$. For a strong guide field of~$b_g = 1.0$, $W_\parallel$ dominates particle injection throughout each simulation.
    
    We also find that larger domains increase the Fermi and pickup injection shares (Figure~\ref{fig:Ninj_vs_time}c, d) and delay injection-share saturation, suggesting that the time required to achieve the limiting partition of injection mechanisms is controlled by the length of the initial current sheet ($\sim \ell_x$).

    \item \textbf{Cumulative injection shares at~$\boldsymbol{t = \tau_f}$:} Larger spatial and temporal domains significantly increase the contributions of pickup and especially Fermi acceleration to particle injection, with larger domains being necessary to achieve convergence for stronger guide fields (Figure~\ref{fig:injection_contributions}).
    
    Adjusting the guide-field strength from~$b_g = 0.1$ to~$0.3$ drastically reduces the pickup injection share, whereas the Fermi injection share remains roughly constant for~$b_g = 0.1$-$0.3$ but then drops significantly as the guide field is raised to~$b_g = 1.0$. In contrast, the~$W_\parallel$ injection share is increased by strengthening the guide field. For a weak guide field of~$b_g = 0.1$ (and the largest domain size), the injection partition by each mechanism is as follows: $\sim 20\%$ of particles are injected by~$W_\parallel$, $\sim 45\%$ by Fermi, and~$\sim 35\%$ by pickup. For a strong guide field of~$b_g = 1.0$, $\sim 70\%$ of particles are injected by~$W_\parallel$, $\sim 25\%$ by Fermi, and~$\sim 5\%$ by pickup (Figure~\ref{fig:injection_contributions}).
\end{enumerate}

\subsection{Post-injection particle acceleration}
\begin{enumerate}
    \item \textbf{Time evolution of high-energy/post-injection ($\boldsymbol{\gamma \gtrsim \gamma_{\rm inj}}$) particle acceleration:} 
    We denote the quantities~$\langle \Delta W_\parallel \rangle(t)$ and~$\langle \Delta W_\perp \rangle(t)$ as the average particle energy gained by parallel and perpendicular electric fields after injection. Stronger guide fields appear to significantly increase~$\langle \Delta W_\parallel \rangle(t)$ immediately after injection, possibly due to increased residual direct acceleration from larger regions with significant $E_\parallel$ around X-points. However, the slope (i.e., the growth rate) of~$\langle \Delta W_\parallel \rangle(t)$ at later times is relatively unaffected by the guide-field strength. On the other hand, stronger guide fields drastically suppress the growth rate of~$\langle \Delta W_\perp \rangle(t)$, both immediately after injection and in the longer term, as indicated by the flatter slopes in Figure~\ref{W_vs_time_postinj}a.
    
    Increasing the spatial domain~$\ell_x$ appears to ``drag out" the evolution of both~$\langle \Delta W_\parallel \rangle$ and~$\langle \Delta W_\perp \rangle$, suggesting that the timescale of post-injection energy gain is controlled by the length of the initial current sheet~$\propto \ell_x$ (Figure~\ref{W_vs_time_postinj}b).

    \item \textbf{High-energy/post-injection~($\boldsymbol{\gamma \gtrsim \gamma_{\rm inj}}$) particle acceleration at~$\boldsymbol{t = \tau_f}$:} We find that~$\langle \Delta W_\perp \rangle$ rises and rapidly converges as~$\ell_x$ increases. The contribution of~$\langle \Delta W_\perp \rangle$ to total post-injection particle energization converges to~$\sim 80\%$ for a weak guide field of~$b_g = 0.1$ and~$\sim 35\%$ for a strong guide field of~$b_g = 1.0$ (Figure~\ref{fig:postinj_results}).
\end{enumerate}

The methodology for studying particle injection presented in this paper is novel and lays the foundation for future work on the topic. This includes defining concrete notions of acceleration efficiency, which yield prospects for future analyses on energy conversion. Furthermore, this is one of the first detailed studies into how guide-field strength affects particle injection from relativistic reconnection, which is crucial for astrophysical applications as guide-field strength in astrophysical objects may vary widely. The demonstration of domain-size convergence of several important particle injection and acceleration parameters presented in this paper will permit the extrapolation and application of those results to the spatial and temporal scales of astrophysical systems.

\section{Acknowledgements} \label{sec:7_acknowlegements}

This work was supported in part by the U.S. Department of Energy, Office of Science, Office of Workforce Development for Teachers and Scientists (WDTS) under the Science Undergraduate Laboratory Internships Program (SULI). We gratefully acknowledge the support from Los Alamos National Laboratory (LANL) through its LDRD program and DoE/OFES support, from the National Science Foundation via grants AST 1806084 and AST 1903335, as well as support from NASA via grants 80NSSC20K0545 and 80HQTR21T0103. This research used resources of the National Energy Research Scientific Computing Center (NERSC), a U.S. Department of Energy Office of Science User Facility located at Lawrence Berkeley National Laboratory, operated under Contract No. DE-AC02-05CH11231. Additional simulations were performed at the Texas Advanced Computing Center (TACC) at The University of Texas at Austin and LANL Institutional Computing Resource.

\appendix
\section{Fitting procedure for a nonthermal spectrum} \label{sec:8_appendix_A}

We fit the nonthermal component of each downstream spectrum~$f_{\rm ds}(\gamma)$ primarily using the fitting procedure described by \citet{Werner2018}. One crucial difference, however, is that we introduce another step in the procedure, which allows us to calculate injection energies.

At the beginning of this procedure, we implement a pool-adjacent-violators smoothing algorithm \citep{Leeuw2009}, which requires~$f_{\rm ds}(\gamma)$ to be monotone for~$\gamma > \gamma_{\rm mono}$ (i.e., the restricted spectrum~$f_{\rm ds}\big|_{\gamma > \gamma_{\rm mono}}$ is a valid input for the algorithm). Different values of~$\gamma_{\rm mono}$ may capture different valid power-law segments (later we clarify what a ``valid" segment is). Therefore, to minimize bias, we prepare a scan over several~$\gamma_{\rm mono}$. We calculate where each power-law segment begins and ends by introducing a ``power-law tolerance" that controls how much $p_\gamma \equiv -\,d \log{f_{\rm ds}(\gamma)}/d \log{\gamma}$ (i.e., the local logarithmic slope of~$f_{\rm ds}(\gamma)$) is allowed to vary from a central value. Similar to~$\gamma_{\rm mono}$, different $p$-tolerances capture different power-law segments, so we also prepare a scan over $p$-tolerances. For~$p$-tolerances, we choose~$\pm 0.10, \pm 0.11, \dots \pm 0.30$ for every simulation. Unlike the scan over~$p$-tolerances, the~$\gamma_{\rm mono}$ scans are customized for each simulation (via trial-and-error\footnote{Values of $\gamma_{\rm mono}$ that are too low yield energy segments in the thermal component, whereas $\gamma_{\rm mono}$ values that are too large yield energy segments in the high-energy cutoff. Examples of typical $\gamma_{\rm mono}$ scans that could properly identify the power-law range were $\gamma_{\rm mono} \in [4, 5, \dots, 10]$ and $\gamma_{\rm mono} \in [10, 11, \dots, 16]$.}), as each power-law spectrum has a different dynamic range.

\begin{figure}[htp!]
    \centering
    \includegraphics[width = 8cm]{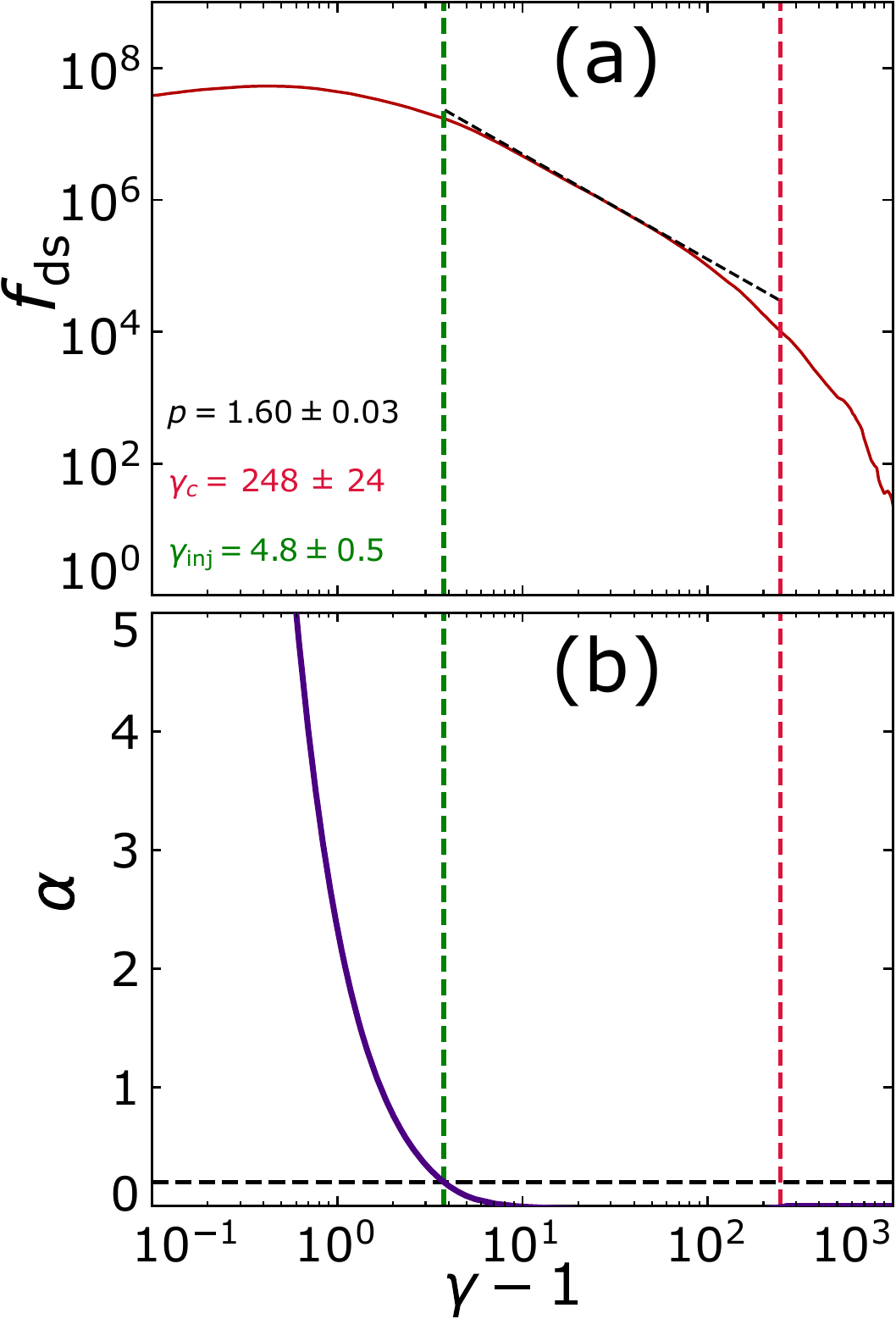}
    \caption{(a): A downstream electron spectrum at the final time~$t = \tau_f$ ($b_g = 0.1$, $\ell_x = 101.8$). The vertical dashed red line shows the high-energy cutoff~$\gamma_c$, which is computed before~$\gamma_{\rm inj}$ (the vertical dashed green line). (b): The low-energy slope correction~$\alpha(\gamma)$ as calculated numerically as a function of~$\gamma$ from Eq.~(\ref{equation:alpha}). The horizontal dashed black line is at~$\alpha' = 0.2$ and defines the threshold for the injection energy. The injection energy is calculated implicitly according to~$\alpha(\gamma_{\rm inj}) = \alpha'$.}
    \label{fig:alpha_parameter}
\end{figure}

In the following sub-appendices, quantities calculated for each unique ($\gamma_{\rm mono}$, $p$-tolerance) pair will have the superscript~``$*$". In particular, we calculate~$p^*$, $\gamma_{c}^*$, and~$\gamma_{\rm inj}^*$. After collecting every starred quantity, we then take (equally weighed) averages of each collection to obtain three characteristic parameters of the nonthermal spectrum: the power-law index~$p$, the high-energy cutoff~$\gamma_{c}$, and the low-energy cutoff (i.e., injection energy)~$\gamma_{\rm inj}$. Note that this fitting procedure may be applied at any time step, as long as monotonicity is satisfied for~$\gamma > \gamma_{\rm mono}$.

\subsection{Calculating \texorpdfstring{$p^*$ and~$\gamma_{c}^*$}{p* and gamma c*}}
For each $p$-tolerance, we define~$p^*$ as the median value of~$p_\gamma(\gamma)$ along the longest logarithmic energy segment for which the variation of~$p_\gamma$ does not exceed the tolerance. This longest segment is found by exhaustion. We then define the high-energy cutoff~$\gamma_{c}^*$ of the power law as the energy at which the downstream spectrum $f_{\rm ds}(\gamma)$ falls below the power-law fit by a factor of~$e$ \citep{Werner2018}.\footnote{The power-law fit is~$f_{\rm pl}^*(\gamma) \equiv A^* \gamma^{-p^*}$, where the prefactor~$A^*$ is normalized so that the fit coincides with the spectrum at the minimum energy of the longest segment.}

\subsection{Calculating \texorpdfstring{$\gamma_{\rm inj}^*$}{gamma inj*}}
A nonthermal spectrum of the form~$f_{\rm nt}^*(\gamma) = f_{\rm pl}^*(\gamma)f_{\rm he}^*(\gamma) = A^*\gamma^{-p^*} e^{-\gamma/\gamma_{c}^*}$ decays at high energies due to the last factor, which accounts for the high-energy cutoff. However, there is no such ``natural" decay at low energies. Therefore, the usual way to define the support of a nonthermal spectrum is~$[\gamma_{\rm inj}^*, \infty)$, where~$\gamma_{\rm inj}^*$ is a (somewhat arbitrarily chosen) value that prevents the power law from reaching low energies (i.e., where the particle spectrum is better described by a Maxwellian distribution). This is the motivation for the current procedure; we seek some ``low-energy cutoff" function that will allow the support of the nonthermal term to be defined appropriately and without bias.

Let us now adopt the Ansatz~$f_{\rm ds} = f_{\rm le}^*f_{\rm pl}^*f_{\rm he}^*$, where~$f_{\rm ds}$ is the downstream spectrum produced by the simulation and~$f_{\rm le}^*$ is the low-energy decay function. We stipulate that~$f_{\rm le}^*(\gamma) = \gamma^{-\alpha^*}$, where~$\alpha^* = \alpha^*(\gamma)$. This is done so that~$\alpha^* \approx 0$ corresponds to the main nonthermal power-law segment, and the extent to which~$\alpha^*$ deviates from~$0$ measures how far the spectrum has deviated from the power law. Solving for~$\alpha^*$, we obtain
\begin{equation} \label{equation:alpha}
    \alpha^*(\gamma) = \log_{\gamma} \bigg(\frac{f_{\rm pl}^*f_{\rm he}^*}{f_{\rm ds}} \bigg) = \log_{\gamma} \bigg(\frac{A^*}{f_{\rm ds}(\gamma)} e^{-\gamma/\gamma_c^*} \bigg) - p^*,
\end{equation}
where~$\alpha^*(\gamma)$ is defined for~$\gamma \in [1, \infty)$.

Figure~\ref{fig:alpha_parameter} shows that~$\alpha$ decays rapidly towards zero as~$\gamma - 1$ approaches the main power-law energy range.\footnote{Note that~$\alpha$ in Figure~\ref{fig:alpha_parameter}, in contrast to~$\alpha^*$, uses Eq.~(\ref{equation:alpha}) but with the corresponding unstarred $A, p, \gamma_c$, i.e., the final obtained values of these quantities.} Therefore we define the injection energy~$\gamma_{\rm inj}^*$ by choosing a standardized threshold~$\alpha' \gtrsim 0$ so that~$\alpha^*(\gamma_{\rm inj}^* - 1) = \alpha'$. This is equivalent to finding the energy~$\gamma_{\rm inj}^*$ at which the local logarithmic slope~$p_\gamma$ deviates from~$p^*$ by~$\alpha'$, i.e., $p^* - p_\gamma = \alpha' = \alpha^*(\gamma_{\rm inj}^* - 1)$. This is a transcendental function that takes~$\alpha'$, $\gamma_c^*$, and~$p^*$ as inputs and returns~$\gamma_{\rm inj}^*$. We chose this threshold to be~$\alpha' = 0.2$ for all ($\gamma_{\rm mono}$, $p$-tolerance) pairs and all simulations. When testing neighboring values of~$\alpha'$, we find that $\gamma_{\rm inj}$ scales roughly as~$1/\sqrt{\alpha'}$, indicating that~$\gamma_{\rm inj}$ is not very sensitive to the choice of~$\alpha'$. In general, $\alpha'$ should be sufficiently small to capture the low-energy cutoff but not too small so as to be accidentally triggered within the error of the power law itself. Accordingly, all of the errors on power-law indices are well below~$0.2$.

We then compute the power-law extent, defined by~$R^* \equiv \gamma_{c}^*/\gamma_{\rm inj}^*$. If~$R^*$ does not exceed some threshold (e.g., we chose 10), all the results for this tolerance and~$\gamma_{\rm mono}$ (i.e., $p^*, \gamma_c^*, \gamma_{\rm inj}^*$) are discarded. In other words, if the ``power law" does not extend beyond one decade, we do not consider it sufficiently well-defined. On the other hand, if~$R^* \geq 10$, the results ($p^*, \gamma_c^*, \gamma_{\rm inj}^*$) are kept.

\subsection{Calculating \texorpdfstring{$p,\gamma_c,\gamma_{\rm inj}$, and~$R$}{p, gamma c, gamma inj, and R}}
We then repeat the above-described process for all tolerances and~$\gamma_{\rm mono}$ within the ranges specified for our scans. Once all quantities ($p^*, \gamma_{\rm inj}^*, \gamma_{c}^*$) are collected for each tolerance and~$\gamma_{\rm mono}$, duplicates (e.g., identical power-law segments resulting from different~$\gamma_{\rm mono}$ or $p$-tolerances) and outliers (e.g., data points beyond~$\pm$ 2 standard deviations from the mean) are removed from each collection.

Finally, ``the" power-law index~$p$, high-energy cutoff~$\gamma_{c}$, and low-energy cutoff (the injection energy)~$\gamma_{\rm inj}$ are set to the (equally-weighed) averages of each collection ($p^*, \gamma_{c}^*$, $\gamma_{\rm inj}^*$). Errors of these quantities are set to one standard deviation of each collection, i.e., containing the central~$\approx 68\%$ of values. This procedure is able to obtain~($p$, $\gamma_{\rm inj}$, $\gamma_c$) with errors around ($0.05$, $0.5$, $30$). The power-law extent~$R$ is calculated after this averaging as the ratio of the finalized~$\gamma_{c}$ and~$\gamma_{\rm inj}$; likewise, the error of~$R$ is calculated by propagating the errors of the finalized~$\gamma_{c}$ and~$\gamma_{\rm inj}$ values.

\vspace{7cm}

\section{Simulation tables} \label{sec:9_appendix_B}

\begin{table}[hb]
\begin{centering}
\begin{tabular}{|r@{\hspace{15pt}}|l|c|c|c|}
    \hline
    \diagbox[innerleftsep=-1.3cm,innerrightsep=-1.1cm]{$\ell_x$}{$b_g$} & $0.1$ & $0.3$ & $0.5$ & $1.0$ \\ 
    \hline
    $72.4$ & 1.57 & 1.83 & 1.90 & 2.20 \\ 
    \hline
    $101.8$ & 1.60 & 1.74 & 1.82 & 2.28 \\ 
    \hline
    $144.8$ & 1.59 & 1.78 & 1.83 & 2.48 \\ 
    \hline
    $203.6$ & 1.72 & 1.82 & 2.06 & 2.70 \\ 
    \hline
    $289.6$ & 1.74 & 1.84 & 2.10 & 2.81 \\ 
    \hline
    $407.3$ & 1.88 & 1.91 & 2.18 & 2.69 \\
    \hline
\end{tabular}
\caption{Power-law indices~$p$ across the parameter scan.}
\label{tab:p}

\vspace{1cm}
\begin{tabular}{|r@{\hspace{15pt}}|l|c|c|c|}
    \hline
    \diagbox[innerleftsep=-1.3cm,innerrightsep=-1.1cm]{$\ell_x$}{$b_g$} & $0.1$ & $0.3$ & $0.5$ & $1.0$ \\ 
    \hline
    $72.4$ & 4.2 & 5.2 & 4.9 & 5.3 \\ 
    \hline
    $101.8$ & 4.8 & 5.3 & 4.7 & 7.9 \\ 
    \hline
    $144.8$ & 4.9 & 5.2 & 5.5 & 10.5 \\ 
    \hline
    $203.6$ & 6.8 & 6.0 & 7.9 & 13.0 \\ 
    \hline
    $289.6$ & 6.9 & 7.0 & 9.1 & 13.8 \\ 
    \hline
    $407.3$ & 9.0 & 7.7 & 9.8 & 14.5 \\
    \hline
\end{tabular}
\caption{Injection energies~$\gamma_{\rm inj}$ across the parameter scan.}
\label{tab:gamma_inj}

\vspace{1cm}
\begin{tabular}{|r@{\hspace{15pt}}|l|c|c|c|}
    \hline
    \diagbox[innerleftsep=-1.3cm,innerrightsep=-1.1cm]{$\ell_x$}{$b_g$} & $0.1$ & $0.3$ & $0.5$ & $1.0$ \\ 
    \hline
    $72.4$ & 242 & 257 & 224 & 182 \\ 
    \hline
    $101.8$ & 248 & 284 & 257 & 189 \\ 
    \hline
    $144.8$ & 308 & 288 & 263 & 242 \\ 
    \hline
    $203.6$ & 416 & 342 & 258 & 259 \\ 
    \hline
    $289.6$ & 505 & 375 & 362 & 330 \\ 
    \hline
    $407.3$ & 739 & 494 & 435 & 251 \\
    \hline
\end{tabular}
\caption{Cutoff energies~$\gamma_c$ across the parameter scan.}
\label{tab:gamma_c}
\end{centering}
\end{table}

\bibliography{main}{}

\begin{thebibliography}{}
\expandafter\ifx\csname natexlab\endcsname\relax\def\natexlab#1{#1}\fi
\providecommand{\url}[1]{\href{#1}{#1}}
\providecommand{\dodoi}[1]{doi:~\href{http://doi.org/#1}{\nolinkurl{#1}}}
\providecommand{\doeprint}[1]{\href{http://ascl.net/#1}{\nolinkurl{http://ascl.net/#1}}}
\providecommand{\doarXiv}[1]{\href{https://arxiv.org/abs/#1}{\nolinkurl{https://arxiv.org/abs/#1}}}

\bibitem[{{Abdo} {et~al.}(2011){Abdo}, {Ackermann}, {Ajello}, {Baldini},
  {Ballet}, {Barbiellini}, {Bastieri}, {Bechtol}, {Bellazzini}, {Berenji},
  {Blandford}, {Bloom}, {Bonamente}, {Borgland}, {Bouvier}, {Bregeon}, {Brez},
  {Brigida}, {Bruel}, {Buehler}, {Buson}, {Caliandro}, {Cameron}, {Cannon},
  {Caraveo}, {Carrigan}, {Casandjian}, {Cavazzuti}, {Cecchi}, {{\c{C}}elik},
  {Charles}, {Chekhtman}, {Chiang}, {Ciprini}, {Claus}, {Cohen-Tanugi},
  {Conrad}, {Cutini}, {de Angelis}, {de Palma}, {Dermer}, {Silva}, {Drell},
  {Dubois}, {Dumora}, {Escande}, {Favuzzi}, {Fegan}, {Finke}, {Focke},
  {Fortin}, {Frailis}, {Fuhrmann}, {Fukazawa}, {Fukuyama}, {Funk}, {Fusco},
  {Gargano}, {Gasparrini}, {Gehrels}, {Georganopoulos}, {Germani}, {Giebels},
  {Giglietto}, {Giommi}, {Giordano}, {Giroletti}, {Glanzman}, {Godfrey},
  {Grenier}, {Guiriec}, {Hadasch}, {Hayashida}, {Hays}, {Horan}, {Hughes},
  {J{\'o}hannesson}, {Johnson}, {Johnson}, {Kadler}, {Kamae}, {Katagiri},
  {Kataoka}, {Kn{\"o}dlseder}, {Kuss}, {Lande}, {Latronico}, {Lee}, {Longo},
  {Loparco}, {Lott}, {Lovellette}, {Lubrano}, {Madejski}, {Makeev},
  {Max-Moerbeck}, {Mazziotta}, {McEnery}, {Mehault}, {Michelson},
  {Mitthumsiri}, {Mizuno}, {Monte}, {Monzani}, {Morselli}, {Moskalenko},
  {Murgia}, {Nakamori}, {Naumann-Godo}, {Nishino}, {Nolan}, {Norris}, {Nuss},
  {Ohsugi}, {Okumura}, {Omodei}, {Orlando}, {Ormes}, {Ozaki}, {Paneque},
  {Panetta}, {Parent}, {Pavlidou}, {Pearson}, {Pelassa}, {Pepe},
  {Pesce-Rollins}, {Pierbattista}, {Piron}, {Porter}, {Rain{\`o}}, {Rando},
  {Razzano}, {Readhead}, {Reimer}, {Reimer}, {Reyes}, {Richards}, {Ritz},
  {Roth}, {Sadrozinski}, {Sanchez}, {Sander}, {Sgr{\`o}}, {Siskind}, {Smith},
  {Spandre}, {Spinelli}, {Stawarz}, {Stevenson}, {Strickman}, {Suson},
  {Takahashi}, {Takahashi}, {Tanaka}, {Thayer}, {Thayer}, {Thompson},
  {Tibaldo}, {Torres}, {Tosti}, {Tramacere}, {Troja}, {Usher}, {Vandenbroucke},
  {Vasileiou}, {Vianello}, {Vilchez}, {Vitale}, {Waite}, {Wang}, {Wehrle},
  {Winer}, {Wood}, {Yang}, {Yatsu}, {Ylinen}, {Zensus}, {Ziegler}, {Fermi LAT
  Collaboration}, {Aleksi{\'c}}, {Antonelli}, {Antoranz}, {Backes}, {Barrio},
  {Becerra Gonz{\'a}lez}, {Bednarek}, {Berdyugin}, {Berger}, {Bernardini},
  {Biland}, {Blanch}, {Bock}, {Boller}, {Bonnoli}, {Bordas}, {Borla Tridon},
  {Bosch-Ramon}, {Bose}, {Braun}, {Bretz}, {Camara}, {Carmona}, {Carosi},
  {Colin}, {Colombo}, {Contreras}, {Cortina}, {Covino}, {Dazzi}, {de Angelis},
  {De Cea del Pozo}, {Delgado Mendez}, {De Lotto}, {De Maria}, {De Sabata},
  {Diago Ortega}, {Doert}, {Dom{\'\i}nguez}, {Dominis Prester}, {Dorner},
  {Doro}, {Elsaesser}, {Ferenc}, {Fonseca}, {Font}, {Garc{\'\i}a L{\'o}pez},
  {Garczarczyk}, {Gaug}, {Giavitto}, {Godinovi}, {Hadasch}, {Herrero},
  {Hildebrand}, {H{\"o}hne-M{\"o}nch}, {Hose}, {Hrupec}, {Jogler}, {Klepser},
  {Kr{\"a}henb{\"u}hl}, {Kranich}, {Krause}, {La Barbera}, {Leonardo},
  {Lindfors}, {Lombardi}, {L{\'o}pez}, {Lorenz}, {Majumdar}, {Makariev},
  {Maneva}, {Mankuzhiyil}, {Mannheim}, {Maraschi}, {Mariotti}, {Mart{\'\i}nez},
  {Mazin}, {Meucci}, {Miranda}, {Mirzoyan}, {Miyamoto}, {Mold{\'o}n},
  {Moralejo}, {Nieto}, {Nilsson}, {Orito}, {Oya}, {Paoletti}, {Paredes},
  {Partini}, {Pasanen}, {Pauss}, {Pegna}, {Perez-Torres}, {Persic}, {Peruzzo},
  {Pochon}, {Prada}, {Prada Moroni}, {Prandini}, {Puchades}, {Puljak},
  {Reichardt}, {Rhode}, {Rib{\'o}}, {Rico}, {Rissi}, {R{\"u}gamer}, {Saggion},
  {Saito}, {Saito}, {Salvati}, {S{\'a}nchez-Conde}, {Satalecka}, {Scalzotto},
  {Scapin}, {Schultz}, {Schweizer}, {Shayduk}, {Shore}, {Sierpowska-Bartosik},
  {Sillanp{\"a}{\"a}}, {Sitarek}, {Sobczynska}, {Spanier}, {Spiro}, {Stamerra},
  {Steinke}, {Storz}, {Strah}, {Struebig}, {Suric}, {Takalo}, {Tavecchio},
  {Temnikov}, {Terzi{\'c}}, {Tescaro}, {Teshima}, {Vankov}, {Wagner},
  {Weitzel}, {Zabalza}, {Zandanel}, {Zanin}, {MAGIC Collaboration}, {Villata},
  {Raiteri}, {Aller}, {Aller}, {Chen}, {Jordan}, {Koptelova}, {Kurtanidze},
  {L{\"a}hteenm{\"a}ki}, {McBreen}, {Larionov}, {Lin}, {Nikolashvili},
  {Reinthal}, {Angelakis}, {Capalbi}, {Carrami{\~n}ana}, {Carrasco}, {Cassaro},
  {Cesarini}, {Falcone}, {Gurwell}, {Hovatta}, {Kovalev}, {Kovalev},
  {Krichbaum}, {Krimm}, {Lister}, {Moody}, {Maccaferri}, {Mori}, {Nestoras},
  {Orlati}, {Pace}, {Pagani}, {Pearson}, {Perri}, {Piner}, {Ros}, {Sadun},
  {Sakamoto}, {Tammi}, \& {Zook}}]{Abdo2011}
{Abdo}, A.~A., {Ackermann}, M., {Ajello}, M., {et~al.} 2011, \apj, 736, 131,
  \dodoi{10.1088/0004-637X/736/2/131}

\bibitem[{{Atoyan}(1999)}]{Atoyan1999}
{Atoyan}, A.~M. 1999, \aap, 346, L49.
\newblock \doarXiv{astro-ph/9905204}

\bibitem[{{Ball} {et~al.}(2019){Ball}, {Sironi}, \& {{\"O}zel}}]{Ball2019}
{Ball}, D., {Sironi}, L., \& {{\"O}zel}, F. 2019, \apj, 884, 57,
  \dodoi{10.3847/1538-4357/ab3f2e}

\bibitem[{Ball {et~al.}(2018)Ball, Sironi, \& Özel}]{Ball2018}
Ball, D., Sironi, L., \& Özel, F. 2018, The Astrophysical Journal, 862, 80,
  \dodoi{10.3847/1538-4357/aac820}

\bibitem[{Beloborodov(2017)}]{Beloborodov2017}
Beloborodov, A.~M. 2017, The Astrophysical Journal, 850, 141,
  \dodoi{10.3847/1538-4357/aa8f4f}

\bibitem[{Bhattacharjee {et~al.}(2009)Bhattacharjee, Huang, Yang, \&
  Rogers}]{Bhattacharjee2009}
Bhattacharjee, A., Huang, Y.-M., Yang, H., \& Rogers, B. 2009, Physics of
  Plasmas, 16, 112102, \dodoi{10.1063/1.3264103}

\bibitem[{Biskamp(2000)}]{Biskamp2000}
Biskamp, D. 2000, Magnetic Reconnection in Plasmas, Cambridge Monographs on
  Plasma Physics (Cambridge University Press), \dodoi{10.1017/CBO9780511599958}

\bibitem[{Blandford {et~al.}(2019)Blandford, Meier, \&
  Readhead}]{Blandford2019}
Blandford, R., Meier, D., \& Readhead, A. 2019, Annual Review of Astronomy and
  Astrophysics, 57, 467–509, \dodoi{10.1146/annurev-astro-081817-051948}

\bibitem[{Bowers {et~al.}(2008)Bowers, Albright, Yin, Bergen, \&
  Kwan}]{Bowers2008}
Bowers, K.~J., Albright, B.~J., Yin, L., Bergen, B., \& Kwan, T. J.~T. 2008,
  Physics of Plasmas, 15, 055703, \dodoi{10.1063/1.2840133}

\bibitem[{Cerutti \& Giacinti(2020)}]{Cerutti2020}
Cerutti, B., \& Giacinti, G. 2020, Astronomy \& Astrophysics, 642, A123,
  \dodoi{10.1051/0004-6361/202038883}

\bibitem[{Cerutti {et~al.}(2015)Cerutti, Philippov, Parfrey, \&
  Spitkovsky}]{Cerutti2015}
Cerutti, B., Philippov, A., Parfrey, K., \& Spitkovsky, A. 2015, Monthly
  Notices of the Royal Astronomical Society, 448, 606,
  \dodoi{10.1093/mnras/stv042}

\bibitem[{Cerutti {et~al.}(2020)Cerutti, Philippov, \& Dubus}]{Cerutti2020b}
Cerutti, B., Philippov, A.~A., \& Dubus, G. 2020, Astronomy \& Astrophysics,
  642, A204, \dodoi{10.1051/0004-6361/202038618}

\bibitem[{Cerutti {et~al.}(2016)Cerutti, Philippov, \&
  Spitkovsky}]{Cerutti2016}
Cerutti, B., Philippov, A.~A., \& Spitkovsky, A. 2016, Monthly Notices of the
  Royal Astronomical Society, 457, 2401, \dodoi{10.1093/mnras/stw124}

\bibitem[{Cerutti {et~al.}(2012)Cerutti, Uzdensky, \& Begelman}]{Cerutti2012}
Cerutti, B., Uzdensky, D.~A., \& Begelman, M.~C. 2012, The Astrophysical
  Journal, 746, 148, \dodoi{10.1088/0004-637x/746/2/148}

\bibitem[{{Cerutti} {et~al.}(2013){Cerutti}, {Werner}, {Uzdensky}, \&
  {Begelman}}]{Cerutti2013}
{Cerutti}, B., {Werner}, G.~R., {Uzdensky}, D.~A., \& {Begelman}, M.~C. 2013,
  \apj, 770, 147, \dodoi{10.1088/0004-637X/770/2/147}

\bibitem[{Cerutti {et~al.}(2014{\natexlab{a}})Cerutti, Werner, Uzdensky, \&
  Begelman}]{Cerutti2014}
Cerutti, B., Werner, G.~R., Uzdensky, D.~A., \& Begelman, M.~C.
  2014{\natexlab{a}}, The Astrophysical Journal, 782, 104,
  \dodoi{10.1088/0004-637x/782/2/104}

\bibitem[{Cerutti {et~al.}(2014{\natexlab{b}})Cerutti, Werner, Uzdensky, \&
  Begelman}]{Cerutti2014b}
---. 2014{\natexlab{b}}, Physics of Plasmas, 21, 056501,
  \dodoi{10.1063/1.4872024}

\bibitem[{Chen {et~al.}(2022)Chen, Uzdensky, \& Dexter}]{Chen2022}
Chen, A.~Y., Uzdensky, D., \& Dexter, J. 2022, Synchrotron Pair Production
  Equilibrium in Relativistic Magnetic Reconnection,  arXiv,
  \dodoi{10.48550/ARXIV.2209.03249}

\bibitem[{{Clausen-Brown} \& {Lyutikov}(2012)}]{Clausen-Brown2012}
{Clausen-Brown}, E., \& {Lyutikov}, M. 2012, \mnras, 426, 1374,
  \dodoi{10.1111/j.1365-2966.2012.21349.x}

\bibitem[{Comisso \& Sironi(2019)}]{Comisso2019}
Comisso, L., \& Sironi, L. 2019, The Astrophysical Journal, 886, 122,
  \dodoi{10.3847/1538-4357/ab4c33}

\bibitem[{Dahlin {et~al.}(2014)Dahlin, Drake, \& Swisdak}]{Dahlin2014}
Dahlin, J.~T., Drake, J.~F., \& Swisdak, M. 2014, Physics of Plasmas, 21,
  092304, \dodoi{10.1063/1.4894484}

\bibitem[{Dahlin {et~al.}(2017)Dahlin, Drake, \& Swisdak}]{Dahlin2017}
---. 2017, Physics of Plasmas, 24, 092110, \dodoi{10.1063/1.4986211}

\bibitem[{{Daughton} {et~al.}(2014){Daughton}, {Nakamura}, {Karimabadi},
  {Roytershteyn}, \& {Loring}}]{Daughton2014}
{Daughton}, W., {Nakamura}, T.~K.~M., {Karimabadi}, H., {Roytershteyn}, V., \&
  {Loring}, B. 2014, Physics of Plasmas, 21, 052307, \dodoi{10.1063/1.4875730}

\bibitem[{de~Leeuw {et~al.}(2009)de~Leeuw, Hornik, \& Mair}]{Leeuw2009}
de~Leeuw, J., Hornik, K., \& Mair, P. 2009, J. Stat. Software, 32(5), 1,
  \dodoi{10.18637/jss.v032.i05}

\bibitem[{Drake {et~al.}(2009)Drake, Cassak, Shay, Swisdak, \&
  Quataert}]{Drake2009}
Drake, J.~F., Cassak, P.~A., Shay, M.~A., Swisdak, M., \& Quataert, E. 2009,
  The Astrophysical Journal, 700, L16, \dodoi{10.1088/0004-637X/700/1/L16}

\bibitem[{Drake {et~al.}(2006)Drake, Swisdak, Che, \& Shay}]{Drake2006}
Drake, J.~F., Swisdak, M., Che, H., \& Shay, M.~A. 2006, Nature, 443, 553,
  \dodoi{doi:10.1038/nature05116}

\bibitem[{Egedal \& Daughton(2013)}]{Egedal2013}
Egedal, A.~L., \& Daughton, W. 2013, Phys. of Plasmas, 20, 061201,
  \dodoi{10.1063/1.4811092}

\bibitem[{Fermi(1949)}]{Fermi1949}
Fermi, E. 1949, Phys. Rev., 75, 1169, \dodoi{10.1103/PhysRev.75.1169}

\bibitem[{Giannios {et~al.}(2009)Giannios, Uzdensky, \&
  Begelman}]{Giannios2009}
Giannios, D., Uzdensky, D.~A., \& Begelman, M.~C. 2009, Monthly Notices of the
  Royal Astronomical Society: Letters, 395, L29–L33,
  \dodoi{10.1111/j.1745-3933.2009.00635.x}

\bibitem[{Giannios {et~al.}(2010)Giannios, Uzdensky, \&
  Begelman}]{Giannios2010}
---. 2010, Monthly Notices of the Royal Astronomical Society, 402, 1649,
  \dodoi{10.1111/j.1365-2966.2009.16045.x}

\bibitem[{Guo {et~al.}(2014)Guo, Li, Daughton, \& Liu}]{Guo2014}
Guo, F., Li, H., Daughton, W., \& Liu, Y.-H. 2014, Physical Review Letters,
  113, \dodoi{10.1103/physrevlett.113.155005}

\bibitem[{Guo {et~al.}(2019)Guo, Li, Daughton, Kilian, Li, Liu, Yan, \&
  Ma}]{Guo2019}
Guo, F., Li, X., Daughton, W., {et~al.} 2019, ApJ, 879, 5,
  \dodoi{10.3847/2041-8213/ab2a15}

\bibitem[{Guo {et~al.}(2021)Guo, Li, Daughton, Li, Kilian, Liu, Zhang, \&
  Zhang}]{Guo2021}
---. 2021, The Astrophysical Journal, 919, 111,
  \dodoi{10.3847/1538-4357/ac0918}

\bibitem[{Guo {et~al.}(2015)Guo, Liu, Daughton, \& Li}]{Guo2015}
Guo, F., Liu, Y.-H., Daughton, W., \& Li, H. 2015, ApJ, 806, 167,
  \dodoi{10.1088/0004-637X/806/2/167}

\bibitem[{{Guo} {et~al.}(2020){Guo}, {Liu}, {Li}, {Li}, {Daughton}, \&
  {Kilian}}]{Guo2020}
{Guo}, F., {Liu}, Y.-H., {Li}, X., {et~al.} 2020, Physics of Plasmas, 27,
  080501, \dodoi{10.1063/5.0012094}

\bibitem[{Guo {et~al.}(2016)Guo, Li, Li, Daughton, Zhang, Lloyd-Ronning, Liu,
  Zhang, \& Deng}]{Guo2016}
Guo, F., Li, X., Li, H., {et~al.} 2016, The Astrophysical Journal, 818, 7,
  \dodoi{10.3847/2041-8205/818/1/L9}

\bibitem[{Guo {et~al.}(2022)Guo, Li, French, Daughton, Matthaeus, Zhang, Liu,
  Kilian, Johnson, \& Li}]{Guo2022_comment}
Guo, F., Li, X., French, O., {et~al.} 2022, Comment on ``Nonideal Fields Solve
  the Injection Problem in Relativistic Reconnection'',  arXiv,
  \dodoi{10.48550/ARXIV.2208.03435}

\bibitem[{Haggerty {et~al.}(2017)Haggerty, Parashar, Matthaeus, Shay, Yang,
  Wan, Wu, \& Servidio}]{Haggerty2017}
Haggerty, C.~C., Parashar, T.~N., Matthaeus, W.~H., {et~al.} 2017, Physics of
  Plasmas, 24, 102308, \dodoi{10.1063/1.5001722}

\bibitem[{Hakobyan {et~al.}(2021)Hakobyan, Petropoulou, Spitkovsky, \&
  Sironi}]{Hakobyan2021}
Hakobyan, H., Petropoulou, M., Spitkovsky, A., \& Sironi, L. 2021, The
  Astrophysical Journal, 912, 48, \dodoi{10.3847/1538-4357/abedac}

\bibitem[{Hakobyan {et~al.}(2019)Hakobyan, Philippov, \&
  Spitkovsky}]{Hakobyan2019}
Hakobyan, H., Philippov, A., \& Spitkovsky, A. 2019, The Astrophysical Journal,
  877, 53, \dodoi{10.3847/1538-4357/ab191b}

\bibitem[{Hakobyan {et~al.}(2022)Hakobyan, Ripperda, \&
  Philippov}]{Hakobyan2022}
Hakobyan, H., Ripperda, B., \& Philippov, A. 2022, Radiative
  reconnection-powered TeV flares from the black hole magnetosphere in M87,
  arXiv, \dodoi{10.48550/ARXIV.2209.02105}

\bibitem[{Hoshino(2022)}]{Hoshino2022}
Hoshino, M. 2022, Physics of Plasmas, 29, 042902, \dodoi{10.1063/5.0086316}

\bibitem[{Hoshino \& Lyubarsky(2012)}]{Hoshino2012}
Hoshino, M., \& Lyubarsky, Y. 2012, Space Sci Rev, 173, 521–533,
  \dodoi{10.1007/s11214-012-9931-z}

\bibitem[{Hoshino {et~al.}(2001)Hoshino, Mukai, Terasawa, \&
  Shinohara}]{Hoshino2001}
Hoshino, M., Mukai, T., Terasawa, T., \& Shinohara, I. 2001, J. Geophys. Res.,
  106, 25979, \dodoi{10.1029/2001ja900052}

\bibitem[{Jaroschek \& Hoshino(2009)}]{Jaroschek2009}
Jaroschek, C.~H., \& Hoshino, M. 2009, Phys. Rev. Lett., 103, 075002,
  \dodoi{10.1103/PhysRevLett.103.075002}

\bibitem[{Jaroschek \& Treumann(2004)}]{Jaroschek2004}
Jaroschek, C.~H., \& Treumann, R.~A. 2004, Physics of Plasmas, 11, 1151,
  \dodoi{10.1063/1.1644814}

\bibitem[{Ji {et~al.}(2022)Ji, Daughton, Jara-Almonte, Le, Stanier, \&
  Yoo}]{Ji2022}
Ji, H., Daughton, W., Jara-Almonte, J., {et~al.} 2022, Nat Rev Phys,
  \dodoi{10.1038/s42254-021-00419-x}

\bibitem[{Johnson {et~al.}(2022)Johnson, Kilian, Guo, \& Li}]{Johnson2022}
Johnson, G., Kilian, P., Guo, F., \& Li, X. 2022, The Astrophysical Journal,
  933, 73, \dodoi{10.3847/1538-4357/ac7143}

\bibitem[{Kagan {et~al.}(2015)Kagan, Sironi, Cerutti, \& Giannios}]{Kagan2015}
Kagan, D., Sironi, L., Cerutti, B., \& Giannios, D. 2015, Space Science
  Reviews, 191, 545, \dodoi{10.1007/s11214-014-0132-9}

\bibitem[{Kilian {et~al.}(2020)Kilian, Li, Guo, \& Zhang}]{Kilian2020}
Kilian, P., Li, X., Guo, F., \& Zhang, Q. 2020, ApJ, 899, 15,
  \dodoi{10.3847/1538-4357/aba1e9}

\bibitem[{Komissarov \& Lyutikov(2011)}]{Komissarov2011}
Komissarov, S.~S., \& Lyutikov, M. 2011, Monthly Notices of the Royal
  Astronomical Society, 414, 2017, \dodoi{10.1111/j.1365-2966.2011.18516.x}

\bibitem[{Kumar \& Zhang(2015)}]{Kumar2015}
Kumar, P., \& Zhang, B. 2015, Physics Reports, 561, 1,
  \dodoi{10.1016/j.physrep.2014.09.008}

\bibitem[{Larrabee {et~al.}(2003)Larrabee, Lovelace, \&
  Romanova}]{Larrabee2003}
Larrabee, D.~A., Lovelace, R. V.~E., \& Romanova, M.~M. 2003, The Astrophysical
  Journal, 586, 72, \dodoi{10.1086/367640}

\bibitem[{Lemoine(2019)}]{Lemoine2019}
Lemoine, M. 2019, Physical Review D, 99, \dodoi{10.1103/physrevd.99.083006}

\bibitem[{Li {et~al.}(2019{\natexlab{a}})Li, Guo, \& Li}]{Li2019a}
Li, X., Guo, F., \& Li, H. 2019{\natexlab{a}}, The Astrophysical Journal, 879,
  12, \dodoi{10.3847/1538-4357/ab223b}

\bibitem[{{Li} {et~al.}(2018{\natexlab{a}}){Li}, {Guo}, {Li}, \&
  {Birn}}]{Li2018a}
{Li}, X., {Guo}, F., {Li}, H., \& {Birn}, J. 2018{\natexlab{a}}, \apj, 855, 80,
  \dodoi{10.3847/1538-4357/aaacd5}

\bibitem[{{Li} {et~al.}(2018{\natexlab{b}}){Li}, {Guo}, {Li}, \&
  {Li}}]{Li2018b}
{Li}, X., {Guo}, F., {Li}, H., \& {Li}, S. 2018{\natexlab{b}}, \apj, 866, 4,
  \dodoi{10.3847/1538-4357/aae07b}

\bibitem[{Li {et~al.}(2019{\natexlab{b}})Li, Guo, Li, Stanier, \&
  Kilian}]{Li2019b}
Li, X., Guo, F., Li, H., Stanier, A., \& Kilian, P. 2019{\natexlab{b}}, The
  Astrophysical Journal, 884, 118, \dodoi{10.3847/1538-4357/ab4268}

\bibitem[{Liu {et~al.}(2015)Liu, Guo, Daughton, Li, \& Hesse}]{Liu2015}
Liu, Y.-H., Guo, F., Daughton, W., Li, H., \& Hesse, M. 2015, Phys. Rev. Lett.,
  114, 095002, \dodoi{10.1103/physrevlett.114.095002}

\bibitem[{Liu {et~al.}(2017)Liu, Hesse, Guo, Daughton, Li, Cassak, \&
  Shay}]{Liu2017}
Liu, Y.-H., Hesse, M., Guo, F., {et~al.} 2017, Phys. Rev. Lett., 118, 085101,
  \dodoi{10.1103/PhysRevLett.118.085101}

\bibitem[{{Liu} {et~al.}(2020){Liu}, {Lin}, {Hesse}, {Guo}, {Li}, {Zhang}, \&
  {Peery}}]{Liu2020}
{Liu}, Y.-H., {Lin}, S.-C., {Hesse}, M., {et~al.} 2020, \apjl, 892, L13,
  \dodoi{10.3847/2041-8213/ab7d3f}

\bibitem[{{Loureiro} {et~al.}(2007){Loureiro}, {Schekochihin}, \&
  {Cowley}}]{Loureiro2007}
{Loureiro}, N.~F., {Schekochihin}, A.~A., \& {Cowley}, S.~C. 2007, Physics of
  Plasmas, 14, 100703, \dodoi{10.1063/1.2783986}

\bibitem[{{Lu} {et~al.}(2021){Lu}, {Guo}, {Kilian}, {Li}, {Huang}, \&
  {Liang}}]{Lu2021}
{Lu}, Y., {Guo}, F., {Kilian}, P., {et~al.} 2021, \apj, 908, 147,
  \dodoi{10.3847/1538-4357/abd406}

\bibitem[{Lyutikov {et~al.}(2018)Lyutikov, Komissarov, Sironi, \&
  Porth}]{Lyutikov2018}
Lyutikov, M., Komissarov, S., Sironi, L., \& Porth, O. 2018, Journal of Plasma
  Physics, 84, 635840201, \dodoi{10.1017/S0022377818000168}

\bibitem[{Majeski {et~al.}(2021)Majeski, Ji, Jara-Almonte, \&
  Yoo}]{Majeski2021}
Majeski, S., Ji, H., Jara-Almonte, J., \& Yoo, J. 2021, Physics of Plasmas, 28,
  092106, \dodoi{10.1063/5.0059017}

\bibitem[{Mehlhaff {et~al.}(2020)Mehlhaff, Werner, Uzdensky, \&
  Begelman}]{Mehlhaff2020}
Mehlhaff, J.~M., Werner, G.~R., Uzdensky, D.~A., \& Begelman, M.~C. 2020,
  Monthly Notices of the Royal Astronomical Society, 498, 799–820,
  \dodoi{10.1093/mnras/staa2346}

\bibitem[{Mehlhaff {et~al.}(2021)Mehlhaff, Werner, Uzdensky, \&
  Begelman}]{Mehlhaff2021}
---. 2021, Monthly Notices of the Royal Astronomical Society, 508, 4532,
  \dodoi{10.1093/mnras/stab2745}

\bibitem[{Melzani {et~al.}(2014)Melzani, Walder, Folini, Winisdoerffer, \&
  Favre}]{Melzani2014}
Melzani, M., Walder, R., Folini, D., Winisdoerffer, C., \& Favre, J.~M. 2014,
  Astronomy \& Astrophysics, 570, A111, \dodoi{10.1051/0004-6361/201424083}

\bibitem[{Mobius {et~al.}(1985)Mobius, Hovestadt, Klecker, Scholer, Gloeckler,
  \& Ipavich}]{Mobius1985}
Mobius, E., Hovestadt, D., Klecker, B., {et~al.} 1985, Nature, 318, 426,
  \dodoi{10.1038/318426a0}

\bibitem[{{Mochol} \& {Petri}(2015)}]{Mochol2015}
{Mochol}, I., \& {Petri}, J. 2015, \mnras, 449, L51,
  \dodoi{10.1093/mnrasl/slv018}

\bibitem[{Nalewajko {et~al.}(2015)Nalewajko, Uzdensky, Cerutti, Werner, \&
  Begelman}]{Nalewajko2015}
Nalewajko, K., Uzdensky, D.~A., Cerutti, B., Werner, G.~R., \& Begelman, M.~C.
  2015, The Astrophysical Journal, 815, 101,
  \dodoi{10.1088/0004-637x/815/2/101}

\bibitem[{Nalewajko {et~al.}(2018)Nalewajko, Yuan, \& Chru{\'{s}
  }li{\'{n}}ska}]{Nalewajko2018}
Nalewajko, K., Yuan, Y., \& Chru{\'{s} }li{\'{n}}ska, M. 2018, Journal of
  Plasma Physics, 84, \dodoi{10.1017/s0022377818000624}

\bibitem[{Nättilä \& Beloborodov(2021)}]{Nattila2021}
Nättilä, J., \& Beloborodov, A.~M. 2021, The Astrophysical Journal, 921, 87,
  \dodoi{10.3847/1538-4357/ac1c76}

\bibitem[{Ortuño-Macías \& Nalewajko(2020)}]{Macias2019}
Ortuño-Macías, J., \& Nalewajko, K. 2020, Monthly Notices of the Royal
  Astronomical Society, 497, 1365, \dodoi{10.1093/mnras/staa1899}

\bibitem[{Petropoulou \& Sironi(2018)}]{Petropoulou2018}
Petropoulou, M., \& Sironi, L. 2018, MNRAS, 481, 5687,
  \dodoi{10.1093/mnras/sty2702}

\bibitem[{Philippov {et~al.}(2019)Philippov, Uzdensky, Spitkovsky, \&
  Cerutti}]{Philippov2019}
Philippov, A., Uzdensky, D.~A., Spitkovsky, A., \& Cerutti, B. 2019, The
  Astrophysical Journal, 876, L6, \dodoi{10.3847/2041-8213/ab1590}

\bibitem[{{Rees} \& {Gunn}(1974)}]{Rees1974}
{Rees}, M.~J., \& {Gunn}, J.~E. 1974, \mnras, 167, 1,
  \dodoi{10.1093/mnras/167.1.1}

\bibitem[{{Romanova} \& {Lovelace}(1992)}]{Romanova1992}
{Romanova}, M.~M., \& {Lovelace}, R.~V.~E. 1992, \aap, 262, 26

\bibitem[{Schoeffler {et~al.}(2019)Schoeffler, Grismayer, Uzdensky, Fonseca, \&
  Silva}]{Schoeffler2019}
Schoeffler, K.~M., Grismayer, T., Uzdensky, D., Fonseca, R.~A., \& Silva, L.~O.
  2019, The Astrophysical Journal, 870, 49, \dodoi{10.3847/1538-4357/aaf1b9}

\bibitem[{Shibata \& Tanuma(2001)}]{Shibata2001}
Shibata, K., \& Tanuma, S. 2001, Earth, Planets and Space, 53, 473,
  \dodoi{10.1186/bf03353258}

\bibitem[{Sironi(2022)}]{Sironi2022}
Sironi, L. 2022, Physical Review Letters, 128,
  \dodoi{10.1103/physrevlett.128.145102}

\bibitem[{Sironi \& Beloborodov(2020)}]{Sironi2020}
Sironi, L., \& Beloborodov, A.~M. 2020, The Astrophysical Journal, 899, 52,
  \dodoi{10.3847/1538-4357/aba622}

\bibitem[{Sironi \& Cerutti(2017)}]{Sironi2017}
Sironi, L., \& Cerutti, B. 2017, in Modelling Pulsar Wind Nebulae (Springer
  International Publishing), 247--277, \dodoi{10.1007/978-3-319-63031-1_11}

\bibitem[{Sironi {et~al.}(2016)Sironi, Giannios, \& Petropoulou}]{Sironi2016}
Sironi, L., Giannios, D., \& Petropoulou, M. 2016, Monthly Notices of the Royal
  Astronomical Society, 462, 48–74, \dodoi{10.1093/mnras/stw1620}

\bibitem[{Sironi {et~al.}(2015)Sironi, Petropoulou, \& Giannios}]{Sironi2015}
Sironi, L., Petropoulou, M., \& Giannios, D. 2015, Monthly Notices of the Royal
  Astronomical Society, 450, 183, \dodoi{10.1093/mnras/stv641}

\bibitem[{{Sironi} \& {Spitkovsky}(2011)}]{Sironi2011}
{Sironi}, L., \& {Spitkovsky}, A. 2011, \apj, 741, 39,
  \dodoi{10.1088/0004-637X/741/1/39}

\bibitem[{Sironi \& Spitkovsky(2014)}]{Sironi2014}
Sironi, L., \& Spitkovsky, A. 2014, The Astrophysical Journal, 783, L21,
  \dodoi{10.1088/2041-8205/783/1/l21}

\bibitem[{Speiser(1965)}]{Speiser1965}
Speiser, T.~W. 1965, J. Geophys. Res., 70, 4219 ,
  \dodoi{10.1029/JZ070i017p04219}

\bibitem[{Sridhar {et~al.}(2022)Sridhar, Sironi, \& Beloborodov}]{Sridhar2022}
Sridhar, N., Sironi, L., \& Beloborodov, A.~M. 2022, Monthly Notices of the
  Royal Astronomical Society, \dodoi{10.1093/mnras/stac2730}

\bibitem[{{Tavani} {et~al.}(2011){Tavani}, {Bulgarelli}, {Vittorini},
  {Pellizzoni}, {Striani}, {Caraveo}, {Weisskopf}, {Tennant}, {Pucella},
  {Trois}, {Costa}, {Evangelista}, {Pittori}, {Verrecchia}, {Del Monte},
  {Campana}, {Pilia}, {De Luca}, {Donnarumma}, {Horns}, {Ferrigno}, {Heinke},
  {Trifoglio}, {Gianotti}, {Vercellone}, {Argan}, {Barbiellini}, {Cattaneo},
  {Chen}, {Contessi}, {D'Ammando}, {DeParis}, {Di Cocco}, {Di Persio},
  {Feroci}, {Ferrari}, {Galli}, {Giuliani}, {Giusti}, {Labanti}, {Lapshov},
  {Lazzarotto}, {Lipari}, {Longo}, {Fuschino}, {Marisaldi}, {Mereghetti},
  {Morelli}, {Moretti}, {Morselli}, {Pacciani}, {Perotti}, {Piano}, {Picozza},
  {Prest}, {Rapisarda}, {Rappoldi}, {Rubini}, {Sabatini}, {Soffitta},
  {Vallazza}, {Zambra}, {Zanello}, {Lucarelli}, {Santolamazza}, {Giommi},
  {Salotti}, \& {Bignami}}]{Tavani2011}
{Tavani}, M., {Bulgarelli}, A., {Vittorini}, V., {et~al.} 2011, Science, 331,
  736, \dodoi{10.1126/science.1200083}

\bibitem[{Uzdensky \& Loureiro(2016)}]{Uzdensky2016b}
Uzdensky, D., \& Loureiro, N. 2016, Physical Review Letters, 116,
  \dodoi{10.1103/physrevlett.116.105003}

\bibitem[{Uzdensky(2011)}]{Uzdensky2011}
Uzdensky, D.~A. 2011, Space Science Reviews, 160, 45–71,
  \dodoi{10.1007/s11214-011-9744-5}

\bibitem[{Uzdensky(2016)}]{Uzdensky2016}
---. 2016, in Magnetic Reconnection (Springer International Publishing),
  473--519, \dodoi{10.1007/978-3-319-26432-5_12}

\bibitem[{Uzdensky(2022)}]{Uzdensky2022}
---. 2022, Journal of Plasma Physics, 88, \dodoi{10.1017/s0022377822000046}

\bibitem[{Uzdensky {et~al.}(2011)Uzdensky, Cerutti, \&
  Begelman}]{Uzdensky2011b}
Uzdensky, D.~A., Cerutti, B., \& Begelman, M.~C. 2011, The Astrophysical
  Journal, 737, L40, \dodoi{10.1088/2041-8205/737/2/l40}

\bibitem[{{Uzdensky} {et~al.}(2011){Uzdensky}, {Cerutti}, \&
  {Begelman}}]{Uzdensky_etal-2011}
{Uzdensky}, D.~A., {Cerutti}, B., \& {Begelman}, M.~C. 2011, \apjl, 737, L40,
  \dodoi{10.1088/2041-8205/737/2/L40}

\bibitem[{Uzdensky {et~al.}(2010)Uzdensky, Loureiro, \&
  Schekochihin}]{Uzdensky2010}
Uzdensky, D.~A., Loureiro, N.~F., \& Schekochihin, A.~A. 2010, Physical Review
  Letters, 105, \dodoi{10.1103/physrevlett.105.235002}

\bibitem[{{Werner} {et~al.}(2019){Werner}, {Philippov}, \&
  {Uzdensky}}]{Werner2019}
{Werner}, G.~R., {Philippov}, A.~A., \& {Uzdensky}, D.~A. 2019, \mnras, 482,
  L60, \dodoi{10.1093/mnrasl/sly157}

\bibitem[{Werner \& Uzdensky(2017)}]{Werner2017}
Werner, G.~R., \& Uzdensky, D.~A. 2017, The Astrophysical Journal Letters, 843,
  L27, \dodoi{10.3847/2041-8213/aa7892}

\bibitem[{Werner \& Uzdensky(2021)}]{Werner2021}
---. 2021, Journal of Plasma Physics, 87, \dodoi{10.1017/s0022377821001185}

\bibitem[{Werner {et~al.}(2018)Werner, Uzdensky, Begelman, Cerutti, \&
  Nalewajko}]{Werner2018}
Werner, G.~R., Uzdensky, D.~A., Begelman, M.~C., Cerutti, B., \& Nalewajko, K.
  2018, Monthly Notices of the Royal Astronomical Society, 473, 4840–4861,
  \dodoi{10.1093/mnras/stx2530}

\bibitem[{Werner {et~al.}(2016)Werner, Uzdensky, Cerutti, Nalewajko, \&
  Begelman}]{Werner2016}
Werner, G.~R., Uzdensky, D.~A., Cerutti, B., Nalewajko, K., \& Begelman, M.~C.
  2016, The Astrophysical Journal Letters, 816, L8,
  \dodoi{10.3847/2041-8205/816/1/L8}

\bibitem[{Yamada(2022)}]{Yamada2022}
Yamada, M. 2022, Magnetic Reconnection: A Modern Synthesis of Theory,
  Experiment, and Observations, Princeton Series in Astrophysics (Princeton
  University Press)

\bibitem[{Yamada {et~al.}(2010)Yamada, Kulsrud, \& Ji}]{Yamada2010}
Yamada, M., Kulsrud, R., \& Ji, H. 2010, Rev. Mod. Phys., 82, 603,
  \dodoi{10.1103/RevModPhys.82.603}

\bibitem[{Zenitani \& Hoshino(2001)}]{Zenitani2001}
Zenitani, S., \& Hoshino, M. 2001, The Astrophysical Journal, 562, L63–L66,
  \dodoi{10.1086/337972}

\bibitem[{{Zenitani} \& {Hoshino}(2005)}]{Zenitani2005}
{Zenitani}, S., \& {Hoshino}, M. 2005, \prl, 95, 095001,
  \dodoi{10.1103/PhysRevLett.95.095001}

\bibitem[{Zenitani \& Hoshino(2007)}]{Zenitani2007}
Zenitani, S., \& Hoshino, M. 2007, The Astrophysical Journal, 670, 702,
  \dodoi{10.1086/522226}

\bibitem[{Zenitani \& Hoshino(2008)}]{Zenitani2008}
---. 2008, The Astrophysical Journal, 677, 530, \dodoi{10.1086/528708}

\bibitem[{{Zhang} {et~al.}(2021{\natexlab{a}}){Zhang}, {Li}, {Giannios}, \&
  {Guo}}]{Zhang2021c}
{Zhang}, H., {Li}, X., {Giannios}, D., \& {Guo}, F. 2021{\natexlab{a}}, \apj,
  912, 129, \dodoi{10.3847/1538-4357/abf2be}

\bibitem[{{Zhang} {et~al.}(2020){Zhang}, {Li}, {Giannios}, {Guo}, {Liu}, \&
  {Dong}}]{Zhang2020}
{Zhang}, H., {Li}, X., {Giannios}, D., {et~al.} 2020, \apj, 901, 149,
  \dodoi{10.3847/1538-4357/abb1b0}

\bibitem[{{Zhang} {et~al.}(2022){Zhang}, {Li}, {Giannios}, {Guo}, {Thiersen},
  {B{\"o}ttcher}, {Lewis}, \& {Venters}}]{Zhang2022}
---. 2022, \apj, 924, 90, \dodoi{10.3847/1538-4357/ac3669}

\bibitem[{{Zhang} {et~al.}(2021{\natexlab{b}}){Zhang}, {Sironi}, \&
  {Giannios}}]{Zhang2021b}
{Zhang}, H., {Sironi}, L., \& {Giannios}, D. 2021{\natexlab{b}}, \apj, 922,
  261, \dodoi{10.3847/1538-4357/ac2e08}

\bibitem[{{Zhang} {et~al.}(2019){Zhang}, {Drake}, \& {Swisdak}}]{Zhang2019}
{Zhang}, Q., {Drake}, J.~F., \& {Swisdak}, M. 2019, Physics of Plasmas, 26,
  072115, \dodoi{10.1063/1.5104352}

\bibitem[{{Zhang} {et~al.}(2021{\natexlab{c}}){Zhang}, {Guo}, {Daughton}, {Li},
  \& {Li}}]{Zhang2021a}
{Zhang}, Q., {Guo}, F., {Daughton}, W., {Li}, H., \& {Li}, X.
  2021{\natexlab{c}}, \prl, 127, 185101, \dodoi{10.1103/PhysRevLett.127.185101}

\bibitem[{Zhdankin {et~al.}(2020)Zhdankin, Uzdensky, Werner, \&
  Begelman}]{Zhdankin2020}
Zhdankin, V., Uzdensky, D.~A., Werner, G.~R., \& Begelman, M.~C. 2020, Monthly
  Notices of the Royal Astronomical Society, 493, 603–626,
  \dodoi{10.1093/mnras/staa284}

\bibitem[{Zweibel \& Yamada(2009)}]{Zweibel2009}
Zweibel, E.~G., \& Yamada, M. 2009, Annu. Rev. Astron. Astrophys., 47, 291,
  \dodoi{10.1146/annurev-astro-082708-101726}

\end{thebibliography}
\bibliographystyle{aasjournal}

\end{document}